\documentstyle[psfig]{esub2acm}

\def\ket#1{|{#1}\rangle}
\def\bra#1{\langle{#1}|}
\def\braket#1#2{\langle{#1}|{#2}\rangle}

\def\col#1#2{\left(\begin{array}{c}#1\\#2\end{array}\right)}
\def\tcol#1#2{(#1, #2)^T}

\def\cos{\mathop{\mbox{cos}}}
\def\sin{\mathop{\mbox{sin}}}
\def\dim{\mathop{\mbox{dim}}}
\def\xor{\mathop{\mbox{xor}}}

\def\mod{\mathop{\mbox{mod}}}
\def\log{\mathop{\mbox{log}}}

\def\Qcontrol{
\begin{picture}(4,1.5)(0,0.5)
\put(0,0.75){\line(1,0){4}}
\put(2,0.75){\circle{0.3}}
\put(2,0.6){\line(0,-1){1.85}}
\end{picture}}

\def\Rtoggle{
\begin{picture}(4,1.5)(0,0.5)
\put(0,0.75){\line(1,0){4}}
\put(2,0.75){\makebox(0,0){$\times$}}
\put(2,0.91){\line(0,-1){0.16}}
\end{picture}}

\def\Qpass{
\begin{picture}(4,1.5)(0,0.5)
\put(0,0.75){\line(1,0){4}}
\end{picture}}

\def\Qcross{
\begin{picture}(4,1.5)(0,0.5)
\put(0,0.75){\line(1,0){4}}
\put(2,0.9){\line(0,-1){2.15}}
\end{picture}}

\def\tbd#1{}

\def\lvert{
\begin{picture}(2,2)(0,0)
\put(1,0){\line(0,1){2}}
\end{picture}}

\def\lcross{
\begin{picture}(2,2)(0,0)
\put(-1,2){\line(2,-1){4}}
\put(-1,0){\line(2,1){4}}
\end{picture}}

\def\lup{
\begin{picture}(2,2)(0,0)
\put(-1,0){\line(2,1){4}}
\end{picture}}

\def\ldown{
\begin{picture}(2,2)(0,0)
\put(-1,2){\line(2,-1){4}}
\end{picture}}

{\catcode`\/=\active \catcode`\.=\active \catcode`\-=\active
  \catcode`\@=\active \gdef\url{\tt\catcode`\/=\active \catcode`\.=\active
  \catcode`\-=\active \catcode`\@=\active
  \def/{\discretionary{\char`\/}{}{\char`\/}}%
  \def.{\discretionary{\char`\.}{}{\char`\.}}%
  \def-{\discretionary{\char`\-}{}{\char`\-}}%
  \def@{\discretionary{\char`\@}{}{\char`\@}}}}

\begin{document}

\title{An Introduction to Quantum Computing for Non-Physicists}
\author{Eleanor Rieffel\\FX Palo Alto Labratory \and Wolfgang Polak\\Consultant}
\date{August 14, 1998}
\sponsor{FX Palo Alto Laboratory,
	 3400 Hillview Avenue,
	 Palo Alto, CA 94304}
\begin{abstract}

Richard Feynman's observation that certain quantum mechanical effects cannot
be simulated efficiently on a computer led to speculation that
computation in general could be done more efficiently if it used
these quantum effects. This speculation proved justified when Peter Shor
described a polynomial time quantum algorithm for factoring integers.

In quantum systems, the computational space increases exponentially
with the size of the system which enables exponential parallelism.
This parallelism could lead to exponentially faster quantum algorithms
than possible classically.  The catch is that accessing the results,
which requires measurement, proves tricky and requires new
non-traditional programming techniques.

The aim of this paper is to guide computer scientists through the
barriers that separate quantum computing from conventional computing.
We introduce basic principles of quantum mechanics to explain where
the power of quantum computers comes from and why it is difficult to
harness.  We describe quantum cryptography, teleportation, and dense
coding.  Various approaches to exploiting the power of quantum
parallelism are explained. We conclude with a discussion of quantum
error correction.

\end{abstract}
\begin{bottomstuff}
\begin{authinfo}
\name{Eleanor Rieffel}
\affiliation{FX Palo Alto Laboratory}
\address{3400 Hillview Avenue, Palo Alto, CA 94304}
\name{Wolfgang Polak}
\address{1021 Yorktown Drive, Sunnyvale, CA 94087}
\end{authinfo}
\permission
\end{bottomstuff}
\markboth{E. Rieffel and W. Polak}{Introduction to Quantum Computing}
\maketitle

\section{Introduction}

Richard Feynman observed in the early 1980's \cite{Feynman-82} that 
certain quantum mechanical effects cannot
be simulated efficiently on a classical computer. This observation 
led to speculation that perhaps
computation in general could be done more efficiently if it made
use of these quantum effects. 
But building quantum computers, computational machines 
that use such quantum effects, proved tricky, and
as no one was sure how to use the quantum effects to speed up computation,
the field developed slowly.
It wasn't until 1994, when Peter Shor surprised the world by describing
a polynomial time quantum algorithm for factoring integers 
\cite{Shor-94,Shor-95}, that the field of
quantum computing came into its own.
This discovery prompted a flurry of activity, both among
experimentalists trying to build quantum computers and theoreticians
trying to find other quantum algorithms.
Additional interest in the subject has been
created by the invention of quantum key distribution and, more 
recently, popular press accounts of experimental successes in 
quantum teleportation and the demonstration of a three-bit quantum computer.  

The aim of this paper is to guide computer scientists and other 
non-physicists through the conceptual and notational barriers that
separate quantum computing from conventional computing and to acquaint 
them with this new and exciting field.  It is important for the 
computer science community to understand these new developments since they
may radically change the way we have to think about computation, 
programming, and complexity.

Classically, the time it takes to do certain computations can be
decreased by using parallel processors.  To achieve an exponential
decrease in time requires an exponential increase in the number of
processors, and hence an exponential increase in the amount of 
physical space needed. However, in quantum systems
the amount of parallelism increases exponentially with the
size of the system. Thus, an exponential increase in parallelism 
requires only a linear increase in the amount of physical space needed.
This effect is called quantum parallelism \cite{Deutsch-Jozsa-91}.

There is a catch, and a big catch at that. While a quantum system
can perform massive parallel computation, access to
the results of the computation is restricted. Accessing 
the results is equivalent
to making a measurement, which disturbs the quantum state. This
problem makes the situation, on the face of it, seem even worse
than the classical situation; we can only read the result of one parallel thread, and
because measurement is probabilistic, we cannot even choose which
one we get.

But in the past few years, various people have found clever ways of
finessing the measurement problem to exploit the power of quantum
parallelism.  This sort of manipulation has no classical analog, and
requires non-traditional programming techniques. One technique
manipulates the quantum state so that a common property of all of the
output values such as the symmetry or period of a function can be read
off.  This technique is used in Shor's factorization algorithm.
Another technique transforms the quantum state to increase the
likelihood that output of interest will be read.  Grover's search
algorithm makes use of such an amplification technique.  This paper
describes quantum parallelism in detail, and the techniques currently
known for harnessing its power.

Section 2, following this introduction, explains of the basic 
concepts of quantum mechanics that are important
for quantum computation.  This section cannot give a comprehensive view
of quantum mechanics.  Our aim is to provide
the reader with tools in the form of mathematics and notation with which
to work with the quantum mechanics involved in quantum computation.  
We hope that this paper will equip 
readers well enough that they can
freely explore the theoretical realm of quantum computing.

Section \ref{qubits} defines the quantum bit, or qubit.  Unlike
classical bits, a quantum bit can be put in a superposition state that
encodes both $0$ and $1$.  There is no good classical explanation of
superpositions: a quantum bit representing $0$ and $1$ can neither be
viewed as ``between'' $0$ and $1$ nor can it be viewed as a hidden unknown state
that represents either $0$ or $1$ with a certain probability.  Even
single quantum bits enable interesting applications. 
We describe the use of a single quantum bit for secure key
distribution.

But the real power of quantum computation derives from the exponential
state spaces of multiple quantum bits: just as a single qubit can be
in a superposition of $0$ and $1$, a register of $n$ qubits can be in
a superposition of all $2^n$ possible values. The ``extra'' states that
have no classical analog and lead to the exponential size of the
quantum state space are the entangled states, like the state leading
to the famous EPR\footnote{EPR = Einstein, Podolsky and Rosen}
paradox (see section \ref{epr}).

We discuss the two types of operations a quantum system can
undergo: measurement and quantum state transformations. Most quantum
algorithms involve a sequence of quantum state transformations followed by a 
measurement. For classical computers there are sets of gates that 
are universal in the sense that any classical computation can be 
performed using a sequence of these gates. Similarly, there are sets of 
primitive quantum state transformations, called quantum gates, that are
universal for quantum computation.  Given enough quantum
bits, it is possible to construct a universal quantum Turing 
machine.

Quantum physics puts restrictions on the types of transformations that
can be done. In particular, all quantum state transformations, and therefore
all quantum gates and all quantum computations, must be reversible.
Yet all classical algorithms can be made reversible and can be computed 
on a quantum computer in comparable time. Some common quantum
gates are defined in section \ref{gates}.

Two applications combining quantum gates and entangled states are 
described in section \ref{coding}: teleportation and dense coding.
Teleportation is the transfer of a quantum state from one place to
another through classical channels. That teleportation is possible
is surprising since quantum mechanics tells us that it is not possible
to clone quantum states or even measure them without disturbing the
state. Thus, it is not obvious what information could be sent through
classical channels that could possibly enable the reconstruction of
an unknown quantum state at the other end.
Dense coding, a dual to teleportation, uses a single quantum bit to transmit
two bits of classical information.  Both teleportation and dense 
coding rely on the entangled states described in the EPR experiment.

It is only in section \ref{computer} that we see where an exponential speed-up
over classical computers might come from. 
The input to a quantum computation can
be put in a superposition state that encodes all possible 
input values.  Performing the computation on this initial state 
will result in superposition of all of the corresponding
output values.  Thus, in the same time it takes to compute the output
for a single input state on a classical computer, a quantum computer
can compute the values for all input states. This process is known
as quantum parallelism. However,
measuring the output states will randomly yield only one of the values in
the superposition, and at the same time destroy
all of the other results of the computation.
Section \ref{computer} describes this situation in
detail. Sections \ref{shor} and \ref{search} describe techniques for taking
advantage of quantum parallelism inspite of the severe constraints
imposed by quantum mechanics on what can be measured.

Section \ref{shor} describes the details of Shor's
polynomial time factoring algorithm.  
The fastest known classical factoring algorithm requires exponential time and
it is generally believed that there is no classical
polynomial time factoring algorithm. Shor's is a beautiful algorithm
that takes advantage of quantum parallelism by using a quantum
analog of the Fourier transform.

Lov Grover developed a technique for
searching an unstructured list of $n$ items 
in $O(\sqrt n)$ steps on a quantum computer. Classical computers
can do no better than $O(n)$, so unstructured
search on a quantum computer is provably more efficient than search on a
classical computer. However, the speed-up is only polynomial, not exponential,
and it has been shown that Grover's algorithm is optimal for quantum 
computers. It seems likely that search algorithms that could take advantage
of some problem structure could do better. Tad Hogg, among others,
has explored such possibilities.
We describe various quantum search techniques in section \ref{search}.

It is as yet unknown whether the power of quantum parallelism can
be harnessed for a wide variety of applications. One tantalizing open
question is whether quantum computers can solve NP complete problems
in polynomial time. 

Perhaps the biggest open question is whether useful quantum computers
can be built. There are a number of proposals for building quantum 
computers using ion traps, nuclear magnetic resonance (NMR), optical 
and solid state techniques. All of the current proposals have scaling
problems, so that a breakthrough will be needed to go beyond tens of
qubits to hundreds of qubits. While both optical and solid state 
techniques show promise, NMR and ion trap technologies are the most 
advanced so far.

In an ion trap quantum computer \cite{Cirac-Zoller-95,Steane-96b} a
linear sequence of ions  representing the qubits are confined by 
electric fields.
Lasers are directed at individual ions to perform single bit quantum
gates.  Two-bit operations are realized by using a laser on one qubit
to create an impulse that ripples through a chain of ions to the
second qubit where another laser pulse stops the rippling and
performs the two-bit operation.  The approach requires that the ions be
kept in extreme vacuum and at extremely low temperatures.
 
The NMR approach has the advantage that it will work at room temperature,
and that NMR technology in general is already fairly advanced.
The idea is to use macroscopic amounts of matter and encode a quantum bit in
the average spin state of a large number of nuclei.  The spin 
states can be manipulated by magnetic fields and the average 
spin state can be measured with NMR techniques.  The main
problem with the technique is that it doesn't scale well; the measured
signal scales as $1/2^n$ with the number of qubits $n$. However, a 
recent proposal \cite{Schulman-Vazirani-98} has been made that may overcome this problem. 
NMR computers with three qubits have
been built successfully \cite{Cory98, Vandersypen99, Gershenfeld-Chuang-97,NMR-GHZ}.
This paper will not discuss further the physical and engineering problems of building quantum computers. 

The greatest problem for building quantum computers is decoherence,
the distortion of the quantum state due to interaction with the environment.
For some time it was feared that quantum computers could not be built
because it would be impossible to isolate them sufficiently from the
external environment. The breakthrough came from the algorithmic
rather than the physical side, through the invention of quantum error
correction techniques. Initially people thought quantum error correction
might be impossible because of the impossibility of reliably copying 
unknown quantum states, but it turns out that it is possible to 
design quantum error correcting codes that detect certain kinds of errors and
enable the reconstruction of the exact error-free 
quantum state.  Quantum error
correction is discussed in section \ref{qec}.

Appendices provide background information
on tensor products and continued fractions.

\section{Quantum Mechanics}

Quantum mechanical\index{quantum mechanics} phenomena are difficult 
to understand since most of our everyday experiences are not 
applicable.  This paper cannot provide
a deep understanding of quantum mechanics 
(see ~\cite{Feynman-65, Liboff, GZ97} for expositions of quantum mechanics).  
Instead, we will 
give some feeling as to the nature of quantum mechanics and 
some of the mathematical formalisms needed to work with
quantum mechanics to the extent needed for quantum computing.

Quantum mechanics\index{quantum mechanics} is a theory in the 
mathematical sense: it is
governed by a set of axioms. The consequences of the axioms 
describe the behavior of quantum systems. The axioms lead to 
several apparent paradoxes: in the Compton effect\index{Compton effect}
it appears as if an action precedes its cause;
the EPR\index{EPR} experiment 
makes it appear as if action over a distance faster than the
speed of light is possible. We will discuss the EPR\index{EPR} experiment in detail in
section \ref{epr}.  Verification of most predictions is indirect, and requires 
careful experimental design and specialized equipment. We will begin,
however, with an experiment that requires only readily available
equipment and that will illustrate some of the key aspects of
quantum mechanics needed for quantum computation.


\subsection{Photon Polarization}

Photons are the only particles\index{particles} that we can directly observe.  The following
simple experiment can be performed with minimal equipment: a strong light
source, like a laser pointer, and
three polaroids (polarization filters) that can be picked up 
at any camera supply 
store.  The experiment demonstrates some of the
principles of quantum mechanics\index{quantum mechanics} through photons and their polarization.

\subsubsection{The Experiment}

A beam of light shines on a projection screen. 
Filters $A$, $B$, and $C$ are
polarized horizontally, at $45^o$, and vertically, respectively, and
can be placed so as to intersect the beam of light.

First, insert filter $A$. Assuming the incoming light is randomly
polarized, the intensity of the output will have half of the
intensity of the incoming light. The outgoing photons are now
all horizontally polarized.  

\begin{center}
\mbox{\psfig{file=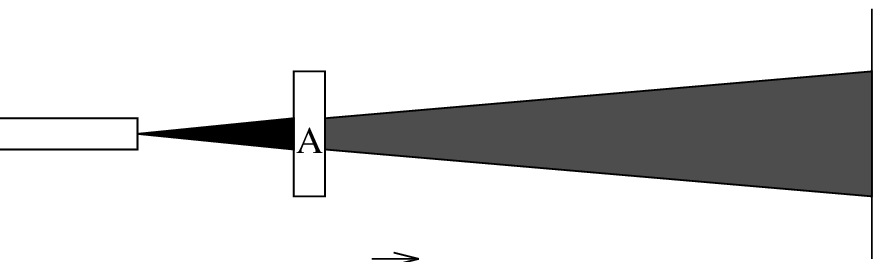,width=3in}}
\end{center}

The function of filter $A$ cannot
be explained as a ``sieve'' that only lets those photons pass that
happen to be already horizontally polarized.  If that were the case,
few of the randomly polarized incoming electrons would be horizontally
polarized, so we would expect a much larger attenuation of the light
as it passes through the filter.

Next, when filter $C$ is inserted the intensity of the output drops to
zero.  None of the horizontally polarized photons can pass
through the vertical filter.  A sieve model could explain this
behavior.

\begin{center}
\mbox{\psfig{file=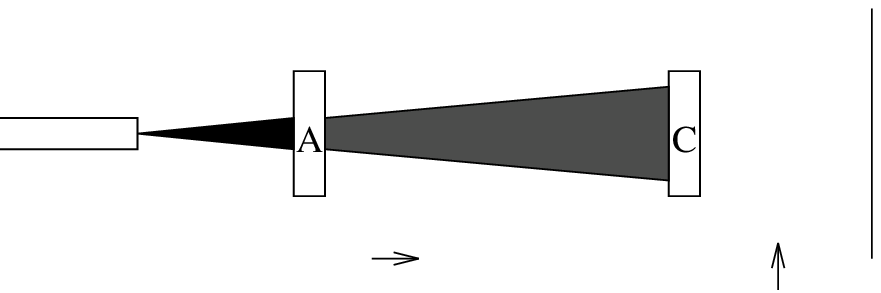,width=3in}}
\end{center}

Finally, after filter $B$ is inserted between $A$ and $C$, a small
amount of light will be visible on the screen, exactly one eighth
of the original amount of light. 

\begin{center}
\mbox{\psfig{file=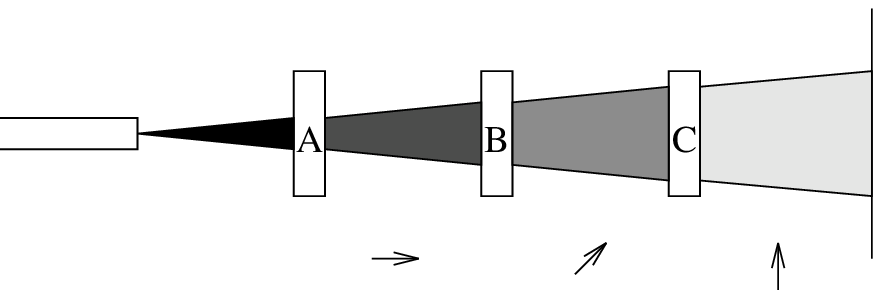,width=3in}}
\end{center}
Here we have a nonintuitive effect.  Classical
experience suggests that adding a filter should only be able to
decrease the number of photons getting through. How can it increase it?

\subsubsection{The Explanation}
\label{explanation}

A photon\index{photon}'s polarization state can be modelled by a unit vector 
pointing in the appropriate direction.  Any arbitrary 
polarization can be expressed as a linear combination
$a \ket{\uparrow} + b \ket{\to}$
of the two basis vectors\footnote{The notation $\ket{\to}$ is 
explained in section \ref{braket}.} $\ket{\to}$ (horizontal
polarization) and $\ket{\uparrow}$ (vertical polarization).

Since we are only interested in the direction of the polarization (the 
notion of ``magnitude'' is not meaningful), the state vector will be
a unit vector, i.e., $\vert a\vert^2 + \vert b\vert^2 = 1$.  
In general, the polarization 
of a photon\index{photon} can be expressed as $a \ket{\uparrow} + b \ket{\to}$ 
where $a$ and $b$ 
are complex numbers\index{complex numbers}\footnote{Imaginary
coefficients correspond to circular polarization.}
such that $\vert a\vert^2 + \vert b\vert^2 = 1$.
Note, the choice of
basis\index{basis} for this representation
is completely arbitrary: any two orthogonal\index{orthogonal} unit vectors\index{unit vectors} will do (e.g. $\{\ket{\nwarrow},\ket{\nearrow}\}$).

\begin{figure}
\begin{center}
\nopagebreak
\mbox{\psfig{file=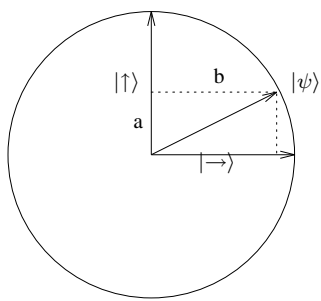,width=1.5in}}\nopagebreak\\
\begin{picture}(0,-1.5)
\put(-1.5, 9){$\ket{\uparrow}$}
\put(1.7,6){$\ket{\to}$}
\put(5.2,9){$\ket{\psi}$}
\end{picture}
\end{center}
\caption{Measurement is a projection onto the basis}
\label{measure-fig}
\end{figure}

The measurement\index{measurement postulate} postulate of quantum
mechanics states that any device  measuring a $2$-dimensional system
has an associated orthonormal basis with respect to which the quantum
measurement takes place. Measurement of a state transforms the state
into one of the measuring device's associated basis vectors.  The
probability that the state is measured as basis vector $\ket{u}$ is
the square of the norm of the amplitude of the component of the
original state in the direction of the basis vector $\ket{u}$.  For
example, given a device for measuring the polarization of photons with
associated basis $\{\ket{\uparrow},\ket{to} \}$, the state $\psi
= a \ket{\uparrow} + b \ket{\to}$ is measured as $\ket{\uparrow}$ with
probability $\vert a\vert^2$ and as $\ket{\to}$ with probability
$\vert b\vert^2$ (see Figure \ref{measure-fig}). Note that different
measuring devices with have different associated basis, and
measurements using these devices will have different outcomes. As
measurements are always made with respect to an orthonormal basis,
throughout the rest of this paper all bases will be assumed to be
orthonormal.

Furthermore, measurement\index{measurement} of the quantum state will change the state to
the result of the measurement.  That is, if measurement of $\psi = a
\ket{\uparrow} + b \ket{\to}$ results in $\ket{\uparrow}$, then the state $\psi$ changes
to $\ket{\uparrow}$ and
a second measurement with respect to the same basis will return
$\ket{\uparrow}$ with probability $1$.  Thus, unless the original state 
happened to be one of the basis vectors, 
measurement will change that state, and 
it is not possible to determine what the original state was.

Quantum mechanics can explain the polarization experiment as follows.
A polaroid measures the quantum state of photons with respect to the basis 
consisting of the vector corresponding to its polarization together
with a vector orthogonal to its
polarization.  The photons which, after being measured by the filter,
match the filter's polarization are let through. The others are
reflected and now have a polarization perpendicular to that of the
filter. For example, filter $A$ measures the photon
polarization\index{photon polarization} with respect to the
basis\index{basis} vector $\ket{\to}$, corresponding to its
polarization.
The photons that pass through filter 
$A$ all have polarization $\ket{\to}$. Those that are reflected by the 
filter all have polarization $\ket{\uparrow}$. 

Assuming that the light source produces photons with random 
polarization, filter $A$ will measure $50\%$ of all 
photons as horizontally polarized.  These photons will
pass through the filter and their state will be 
$\ket{\to}$.  Filter $C$ will
measure these photons with respect to $\ket{\uparrow}$.  But the state
$\ket{\to} = 0 \ket{\uparrow} + 1\ket{\to}$ will be projected onto 
$\ket{\uparrow}$ with probability $0$ and no photons will pass filter $C$.

Finally, filter $B$ measures the quantum state with respect to the basis\index{basis}
\begin{displaymath}
\{{1\over \sqrt 2}(\ket{\uparrow} + \ket{\to}), {1\over \sqrt 2}(\ket{\uparrow} - \ket{\to})\}
\end{displaymath}
 which
we write as $\{\ket{\nearrow}, \ket{\nwarrow}\}$.
Note that $\ket{\to} = {1\over \sqrt 2}(\ket{\nearrow} - \ket{\nwarrow})$
and  $\ket{\uparrow} = {1\over \sqrt 2}(\ket{\nearrow} + \ket{\nwarrow})$.
Those photons that are measured as $\ket{\nearrow}$ pass through the filter.
Photons passing through $A$ with state $\ket{\to}$ will be measured by $B$
as $\ket{\nearrow}$ with probability $1/2$ and so 
$50\%$ of the photons passing through $A$ will pass through $B$ and be in state
$\ket{\nearrow}$.  As before, these photons will
be measured by filter $C$ as $\ket{\uparrow}$ with probability $1/2$.
Thus only one eighth of the original photons manage to pass through 
the sequence of filters $A$, $B$, and $C$.

\subsection{State Spaces and Bra/Ket Notation}
\label{braket}
 
The state space of a quantum system, consisting of the positions, momentums,
polarizations, spins, etc.~of the various particles\index{particles},
is modelled by a Hilbert space of
wave functions. We will not look at the details of these wave functions.
For quantum computing we need only deal with finite quantum
systems and it suffices to consider finite dimensional complex
vector spaces with an inner product that are spanned by abstract
wave functions such as $\ket{\rightarrow}$.  
 
Quantum state spaces and the tranformations acting on them can
be described in terms of vectors and matrices or in the more
compact bra/ket notation invented by Dirac \cite{Dirac-58}\index{Dirac}.  Kets \index{ket} like
$\ket x$ denote column vectors and are
typically used to describe quantum states\index{quantum states}.
The matching bra, $\bra x$, denotes the conjugate 
transpose\index{conjugate transpose} of $\ket x$.
For example, the orthonormal basis $\{\ket 0, \ket 1\}$
can be expressed as $\{\tcol 10, \tcol 01\}$.
Any complex linear combination of  $\ket 0$ and $\ket 1$, $a\ket 0 +b\ket 1$,
can be written  $\tcol ab$.
Note that the choice of the order of the basis vectors is
arbitrary. For example, representing $\ket{0}$ as $\tcol 01$ and
$\ket 1$ as $\tcol 10$ would be fine as long as this is done consistently.
 
Combining $\bra{x}$ and
$\ket{y}$ as in $\bra x\ket y$, also written as $\braket xy$, denotes the inner
product of the two vectors.  For instance, since $\ket 0$ is a unit vector
we have $\braket 00 = 1$ and since $\ket 0$ and $\ket 1$ are orthogonal\index{or
thogonal} we
have $\braket 01 = 0$. 
 
The notation $\ket x\bra y$ is the outer product of $\ket x$ and $\bra
y$.  For example, $\ket 0\bra 1$ is the transformation
that maps $\ket 1$ to $\ket 0$ and $\ket 0$ to $\tcol 00$ since
$$\begin{array}{l}
\ket 0\bra 1\ket 1 = \ket 0\braket 11 = \ket 0\\
\ket 0\bra 1\ket 0 = \ket 0\braket 10 = 0 \ket 0 = \col 00.\\
  \end{array}$$
Equivalently, $\ket 0\bra 1$ can be written in matrix form where
$\ket 0 = \tcol 10$, $\bra 0 = (1, 0)$, $\ket 1 = \tcol 01$, and $\bra 1 = (0, 1
)$.
Then
$$\ket 0\bra 1 = \col 10 (0, 1) = \left(\begin{array}{cc}0&1\\0&0\end{array}\right).$$
 
This notation gives us a convenient way of specifying transformations on
quantum states\index{quantum states} in terms 
of what happens to the basis vectors
(see section \ref{gates}).  For example, the transformation
that exchanges $\ket 0$ and $\ket 1$ is given by the matrix
$$X = \ket 0\bra 1 + \ket 1\bra 0.$$
In this paper we will prefer the slightly more intuitive notation
$$\begin{array}{lrcl}
X:& \ket{0} & \to & \ket{1}\\
  & \ket{1} & \to & \ket{0}\\
  \end{array}
$$
that explicitly specifies the result of a transformation on the basis vectors.
 
\section{Quantum Bits}

\label{qubits}

A quantum bit, or qubit\index{qubit}, is a unit vector in a two dimensional
complex vector space for which a particular basis, denoted by  
$\{\ket 0, \ket 1\}$, has been fixed. The orthonormal basis
$\ket 0$ and $\ket 1$ may correspond to the $\ket{\uparrow}$ and 
$\ket{\to}$ polarizations of a photon respectively, or to the polarizations
$\ket{\nearrow}$ and $\ket{\nwarrow}$. Or $\ket 0$ and $\ket 1$ could 
correspond to the spin-up and spin-down states of an electron.
When talking about qubits, and quantum
computations in general, a fixed basis with respect to which all
statements are made has been chosen in advance. In particular, 
unless otherwise specified, all measurements will be made with 
respect to the standard basis for quantum computation, $\{\ket 0, \ket 1\}$. 

For the purposes of quantum computation, the basis states 
$\ket 0$ and $\ket 1$ are taken to represent the classical bit values
$0$ and $1$ respectively. 
Unlike classical bits\index{classical bits} however, qubits can be in a superposition\index{superposition} of
$\ket 0$ and $\ket 1$ such as $a\ket 0 + b\ket 1$
where $a$ and $b$ are complex numbers such that 
$\vert a\vert^2 + \vert b\vert^2 = 1$. Just as in the 
photon polarization\index{photon polarization} case, if such a superposition\index{superposition} is measured with
respect to the basis $\{\ket 0,\ket 1\}$, the probability that the 
measured value is $\ket 0$ is $\vert a\vert ^2$ and the probability that the
measured value is $\ket 1$ is  $\vert b\vert ^2$.

\label{information}
Even though a quantum bit can be put in infinitely many 
superposition\index{superposition} states, it is only possible to 
extract a single classical bit's worth of information from a single
quantum bit\index{quantum bit}. 
The reason that no more information can be gained from a
qubit\index{qubit} than in a classical bit is that information can only be 
obtained by measurement. When a qubit is measured,
the measurement\index{measurement} 
changes the state to one of the basis states in the way seen 
in the photon polarization\index{photon polarization} experiment. As
every measurement can result in only one of two states, one of the
basis vectors associated to the given measuring device, 
so, just as in the classical case, there are only two possible results.
As measurement changes the state, one cannot measure the state 
of a qubit in
two different bases. Furthermore, as we shall see in the section 
\ref{NoCloning}, quantum states cannot be cloned so it is not possible 
to measure a qubit in two ways, even indirectly by, say, copying the
qubit and measuring the copy in a different basis from the original.

\subsection{Quantum Key Distribution}

Sequences of single qubits can be used to transmit private keys on
insecure channels. 
In 1984 Bennett and Brassard described the first
quantum key distribution scheme \cite{BenBra87,Bennett:1992:QC}. 
Classically, public key encryption techniques, e.g.~RSA, are used for key 
distribution.

Consider the situation in which 
Alice and Bob want to agree on a secret key so that they can 
communicate privately.  
They are connected by an ordinary
bi-directional open channel and a uni-directional quantum channel 
both of which can be observed by Eve, who wishes to eavesdrop on their
conversation. This situation is illustrated in the figure below.
The quantum channel allows Alice to send individual particles (e.g.~photons) to Bob who
can measure their quantum state.  Eve can attempt to measure the state of these
particles and can resend the particles to Bob.

\begin{center}
\mbox{\psfig{file=keysetup.ps,width=4in}}
\end{center}

To begin the process of establishing a secret key,
Alice sends a sequence of bits to Bob by encoding each bit in the
quantum state of a photon as follows.  For each bit, Alice randomly uses
one of the following two bases for encoding each bit:
$$\begin{array}{lrcl}
  & 0 & \to & \ket{\uparrow}\\
  & 1 & \to & \ket{\to}\\
\end{array}
$$
or
$$\begin{array}{lrcl}
  & 0 & \to & \ket{\nwarrow}\\
  & 1 & \to & \ket{\nearrow}.\\
\end{array}
$$
Bob measures the state of the photons he receives by randomly
picking either basis.  After the bits have been transmitted, Bob and Alice
communicate the basis they used for encoding and decoding of each bit
over the open channel.  With this information both can determine which
bits have been transmitted correctly, by identifying those bits 
for which the sending and
receiving bases agree.  They will use these
bits as the key and discard all the others.  On average, Alice and Bob 
will agree on $50\%$ of all bits transmitted.

Suppose that Eve measures the state of the photons transmitted by Alice 
and resends new photons with the measured state.  In this process she 
will use the wrong basis\index{basis} approximately $50\%$ of 
the time, in which case
she will resend the bit with the wrong basis.  So when
Bob measures a resent qubit with the correct basis 
there will be a $25\%$ probability that he measures the 
wrong value.  Thus any eavesdropper on the quantum channel is 
bound to introduce a high error rate that Alice and Bob can detect
by communicating a sufficient number of parity bits of their 
keys over the open channel. So, not only is it likely that
Eve's version of the key is $25\%$ incorrect, but the fact that
someone is eavesdropping\index{eavesdropping} will be apparent to 
Alice and Bob.

Other techniques for exploiting quantum effects for key distribution
have been proposed. See, for example, Ekert\index{Ekert}
\cite{ERTP92}, Bennett\index{Bennet} \cite{Bennet92} and Lo and Chau
\cite{Lo-Chau}.  But none of the quantum key distribution techniques
are substitutes for public key encryption schemes.  Attacks by
eavesdroppers other than the one described here are possible.
Security against all such schemes are discussed in both Mayers
\cite{Mayers} and Lo and Chau \cite{Lo-Chau}.

Quantum key distribution\index{quantum key distribution} has been realized over a distance of 24 km using
standard fiber optical cables \cite{Hughes-etal97} and over 0.5 km 
through the atmosphere \cite{Hughes99}.

\subsection{Multiple Qubits}

Imagine a macroscopic physical 
object breaking apart and multiple pieces flying
off in different directions. The state of this system can be described
completely by describing the state of each of its component pieces separately.
A surprising and unintuitive aspect of the state space of an $n$
particle quantum system is that the state of the system cannot always be
described in terms of the state of its component pieces. 
It is when examining systems of more than one qubit that one first gets
a glimpse of where the computational power of quantum computers could
come from. 

As we saw, the state of a qubit can be represented by a vector in the two
dimensional complex vector space spanned by $\ket 0$ and $\ket 1$. In
classical physics, the possible states of a system of $n$ particles, whose
individual states can be described by a vector in a two dimensional
vector space, form a vector space of $2n$ dimensions. However, in a 
quantum system the resulting state space is much larger; a system of
$n$ qubits has a state space of $2^n$ dimensions.\footnote{ Actually, as we
shall see, the state space is the set of normalized vectors in this
$2^n$ dimensional space, just as the state $a\ket 0+b\ket 1$ of a qubit
is normalized so that $|a|^2 + |b|^2 =1$.} It is this exponential
growth of the state space with the number of particles that suggests
a possible exponential speed-up of computation on quantum computers over
classical computers.

Individual state spaces of $n$ particles
combine classically through the cartesian product. Quantum states,
however, combine through the tensor product\index{tensor product}. Details on properties
of tensor products and their expression in terms of vectors and 
matrices is given in Appendix \ref{tensor-product}.  Let us look briefly at
distinctions between the cartesian product and the tensor product 
that will be crucial to understanding quantum computation. 

Let $V$ and $W$ be two 2-dimensional complex vector spaces with bases
$\{v_1, v_2\}$ and $\{w_1, w_2\}$ respectively. The cartesian product of
these two spaces can take as its basis the union of the bases of its component
spaces $\{v_1, v_2, w_1, w_2\}$. Note that the order
of the basis was chosen arbitrarily. In particular, the dimension 
of the state space of multiple classical
particles grows linearly with the number of particles, 
since  $\dim(X\times Y) = \dim(X) + \dim(Y)$.
The tensor product\index{tensor product} of $V$ and $W$
has basis  
$\{v_1\otimes w_1, v_1\otimes w_2, v_2\otimes w_1, v_2\otimes w_2\}$.
Note that the order of the basis, again, is arbitrary\footnote{It is only when we use matrix 
notation to describe state transformations that the order of basis vectors
becomes relevant.}. So the state space
for two qubits, each with basis $\{\ket 0, \ket 1\}$, has basis 
$\{\ket 0\otimes\ket 0, \ket 0\otimes\ket 1, \ket 1\otimes\ket 0, \ket 1\otimes\ket 1\}$
which can be written more compactly as 
$\{\ket{00}, \ket{01}, \ket{10}, \ket{11}\}$.  More generally, we write 
$\ket x$ to mean $\ket{b_nb_{n-1}\dots b_0}$ where $b_i$ are the binary digits
of the number $x$.

A basis\index{basis} for a three qubit system is
\begin{displaymath}
\{\ket{000},\ket{001},\ket{010},\ket{011},\ket{100},\ket{101},\ket{110},\ket{111}\}
\end{displaymath}
and in general an $n$ qubit system has $2^n$ basis vectors. 
We can now see the exponential growth of the state space with the number
of quantum particles. 
The tensor product\index{tensor product} $X\otimes Y$ has dimension $dim(X)\times dim(Y)$.

The state $\ket{00}+\ket{11}$ is an example of a quantum state that cannot be
described in terms of the state of each of its components (qubits) separately.
In other words, we cannot find 
$a_1,a_2,b_1,b_2$ such that 
$(a_1\ket 0 + b_1\ket 1)\otimes (a_2\ket 0 + b_2\ket 1) = \ket{00}+\ket{11}$ since 
\begin{displaymath}
(a_1\ket 0 + b_1\ket 1)\otimes (a_2\ket 0 + b_2\ket 1) = 
  a_1a_2\ket{00} + a_1b_2\ket{01} + b_1a_2\ket{10} + b_1b_2\ket{11}
\end{displaymath}
 and
$a_1b_2 = 0$ implies that either $a_1a_2 = 0$ or $b_1b_2 = 0$.
States which cannot be decomposed in this way are called entangled\index{entangled} states.
These states represent situations that have no classical counterpart, and
for which we have no intuition. These are also the states that provide
the exponential growth of quantum state spaces with the number of
particles.

Note that it would require vast resources to simulate even a small
quantum system on traditional computers. 
The evolution\index{evolution} of quantum systems is exponentially faster than
their classical simulations\index{simulations}.
The reason for the potential power of quantum computer\index{quantum computer}s is the
possibility of exploiting the quantum state evolution as 
a computational mechanism.

\subsection{Measurement}

The experiment in section \ref{explanation} illustrates how measurement 
of a single qubit projects
the quantum state on to one of the basis states associated with
the measuring device.  The result of a measurement
is probabilistic and the process of measurement changes the 
state to that measured.

Let us look at an example of measurement in a two qubit system. Any 
two qubit state can be expressed as 
$a\ket{00}+b\ket{01}+c\ket{10}+d\ket{11}$, where $a$, $b$, $c$ and
$d$ are complex numbers\index{complex numbers} such that 
$|a|^2+|b|^2+|c|^2+|d|^2 = 1$.  Suppose we wish to measure the 
first qubit with respect to the standard basis
$\{\ket{0}, \ket{1}\}$.  For convenience we will rewrite the state
as follows: 
\begin{eqnarray*}
\lefteqn{a\ket{00}+b\ket{01}+c\ket{10}+d\ket{11}}\\
 & = & \ket{0} \otimes (a\ket{0}+b\ket{1}) + 
       \ket{1} \otimes (c\ket{0}+d\ket{1})\\
 & = & u \ket{0} \otimes (a/u \ket{0}+ b/u \ket{1}) + \\ 
&&    v \ket{1} \otimes (c/v \ket{0}+ d/v \ket{1}).\\
\end{eqnarray*}
For $u = \sqrt{|a|^2+|b|^2}$ and $v = \sqrt{|c|^2+|d|^2}$
the vectors $a/u \ket{0}+ b/u \ket{1}$ and $c/v \ket{0}+ d/v \ket{1}$ are of unit
length.  Once the state has been rewritten as above, as 
a tensor product of the bit
being measured and a second vector of unit length, 
the probabalistic result of a measurement is easy to read off.
Measurement of the first bit will with probability $u^2 = {|a|^2+|b|^2}$
return $\ket{0}$ projecting the state to 
$\ket{0}\otimes(a/u \ket{0}+b/u \ket{1})$ or
with probability $v = {|c|^2+|d|^2}$ yield $\ket{1}$ 
projecting the state to 
$\ket{1} \otimes (c/v \ket{0}+ d/v \ket{1})$.
As $\ket{0}\otimes(a/u \ket{0}+b/u \ket{1})$ and 
$\ket{1} \otimes (c/v \ket{0}+ d/v \ket{1})$ are both unit vectors,
no scaling is necessary.
Measuring the second bit works similarly.  

For the purposes of quantum computation, multi-bit measurement can be
treated as a series of single-bit measurements in the standard basis.
Other sorts of measurements are possible, like measuring whether two
qubits have the same value without learning the actual value of the
two qubits. But such measurements are equivalent to unitary
transformations followed by a standard measurement of individual
qubits, and so it suffices to look only at standard measurements.  

In the two qubit example, the state space is a cartesian product of
the subspace consisting of all states whose first qubit is in the
state $\ket 0$ and the orthogonal subspace of states whose first qubit
is in the state $\ket 1$. Any quantum state can be written as the sum
of two vectors, one in each of the subspaces.  A measurement of $k$
qubits in the standard basis has $2^k$ possible outcomes $m_i$.  Any
device measuring $k$ qubits of an $n$-qubit system splits of the
$2^n$-dimensional state space $\cal H$ into a cartesian product of
orthogonal subspaces $S_1, \dots, S_{2^k}$ with ${\cal H} = S_1
\times\dots\times S_{2^k}$, such that the value of the $k$ qubits
being measured is $m_i$ and the state after measurement is in space
the space $S_i$ for some $i$.  The device randomly chooses one of the
$S_i$'s with probability the square of the amplitude of the component
of $\psi$ in $S_i$, and projects the state into that component,
scaling to give length $1$.  Equivalently, the probability that the
result of the measurement\index{measurement} is a given value is the
sum of the squares of the the absolute values of the amplitudes of all
basis\index{basis} vectors compatible with that value of the
measurement\index{measurement}.

Measurement gives another way of thinking about entangled
particles. Particles are not entangled if the measurement\index{measurement} of one
has no effect on the other. For instance, the state 
$\frac{1}{\sqrt{2}}(\ket{00}+\ket{11})$ is entangled\index{entangled} since the 
probability that the first bit is measured to be $\ket 0$ is $1/2$ 
if the second bit has not been measured. However, if the second bit
had been measured, the probability that the first bit is 
measured as $\ket 0$ is either $1$ or $0$, depending on whether the
second bit was measured as $\ket 0$ or $\ket 1$ respectively. Thus
the probable result of measuring the first bit is changed by a
measurement\index{measurement} of the second bit. On the other hand, the state
$\frac{1}{\sqrt{2}}(\ket{00}+\ket{01})$ is not entangled\index{entangled}: since 
$\frac{1}{\sqrt{2}}(\ket{00}+\ket{01}) = \ket 0\otimes \frac{1}{\sqrt{2}}(\ket{0}+\ket{1})$, any 
measurement\index{measurement} of the first bit will yield $\ket 0$ regardless of
whether the second bit was measured.  Similarly, the second bit has a 
fifty-fifty chance of being measured as $\ket 0$ regardless of 
whether the first bit was measured or not. Note that entanglement, in the
sense that measurement of one particle has an effect on measurements of
another particle, is equivalent to our previous definition of entangled
states as states that cannot be written as a tensor product of individual
states.

\subsection{The EPR Paradox}
 
\label{epr}
Einstein\index{Einstein}, Podolsky\index{Podolsky} and
Rosen\index{Rosen} proposed a gedanken experiment that uses entangled
particles in a manner that seemed to violate fundamental principles
relativity.  Imagine a source that generates two maximally entangled
particles ${1\over \sqrt 2} \ket {00} + {1\over \sqrt 2} \ket {11}$,
called an EPR\index{EPR} pair, and sends one each to Alice and Bob.
 
\begin{center}
\mbox{\psfig{file=epr.ps,width=4in}}
\end{center}

Alice and Bob can be arbitrarily far apart.  Suppose that Alice 
measures her particle and observes state $\ket 0$.  This 
means that the combined state will now be
$\ket {00}$ and if now Bob measures his particle he will 
also observe $\ket 0$.  Similarly, if Alice measures $\ket 1$, 
so will Bob.  Note that the change
of the combined quantum state occurs instantaneously 
even though the two particles
may be arbitrarily far apart.  It appears that this
would enable Alice and Bob to communicate faster than the speed of light.
Further analysis, as we shall see, shows that  even though
there is a coupling between the two particles, there is no way for Alice or
Bob to use this mechanism to communicate.

There are two standard ways that people use to describe entangled\index{entangled}
states and their measurement\index{measurement}. Both have their positive aspects, but
both are incorrect and can lead to misunderstandings. Let us 
examine both in turn.

Einstein\index{Einstein},  Podolsky\index{Podolsky} and Rosen\index{Rosen}
proposed that each particle has some
internal state that completely determines what
the result of any given measurement will be. This state is, for
the moment, hidden from us, and therefore the best we can currently
do is to give probabilistic predictions. Such a theory is known as
a local hidden variable theory. The simplest hidden variable theory
for an EPR\index{EPR} pair is that the particles are either both in state $\ket 0$
or both in state $\ket 1$, we just don't happen to know which. In such
a theory no communication between possibly distant particles is 
necessary to explain the correlated measurements. However, this
point of view cannot explain the results of measurements with
respect to a different basis\index{basis}. In fact, Bell\index{Bell} showed that any local hidden
variable theory predicts that certain measurements will
satisfy an inequality, known as Bell's inequality. However, the result
of actual experiments performing these measurements show that Bell's
inequality is violated. Thus quantum mechanics cannot be explained by any
local hidden variable theory. See \cite{GZ97} for a highly readable
account of Bell's theorem and related experiments.

The second standard description is in terms of cause and effect. 
For example,
we said earlier that a measurement performed by Alice affects a
measurement performed by Bob. However, this view is incorrect also,
and results, as Einstein\index{Einstein}, Podolsky\index{Podolsky} and
Rosen\index{Rosen} recognized, in deep inconsistencies when combined
with relativity theory. It is possible to set up the EPR\index{EPR}
scenario so that one observer sees Alice measure first, then Bob,
while another observer sees Bob measure first, then Alice. According
to relativity, physics must equally well explain the observations of
the first observer as the second. While our terminology of cause and
effect cannot be compatible with both observers, the actual
experimental values are invariant under change of observer. The experimental results
can be explained equally well by Bob's measuring first and causing a
change in the state of Alice's particle, as the other way around. This
symmetry shows that Alice and Bob cannot, in fact, use their
EPR\index{EPR} pair to communicate faster than the speed of light, and
thus resolves the apparent paradox\index{paradox}.  All that can be
said is that Alice and Bob will observe the same random behavior.

As we will see in the section on dense coding\index{dense coding} and
teleportation\index{teleportation}, EPR pairs\index{EPR} can be used to aid
communication, albeit communication slower than the speed of light.

\section {Quantum Gates} \label{gates}

So far we have looked at static quantum systems which change only when
measured. The dynamics of a quantum system, when not being measured, are
governed by Schr\"odinger's equation; the dynamics must take states to 
states in a way that preserves orthogonality. For a complex vector space,
linear transformations that preserve orthogonality are unitary transformations,
defined as follows.  Any linear transformation on a complex vector space
can be described by a matrix.  Let $M^*$ denote the conjugate transpose of the matrix
$M$.  A matrix $M$ is unitary (describes a unitary transformation) if $MM^* = I$. 
Any unitary transformation
of a quantum state space is a legitimate quantum transformation, and vice
versa. One can think of unitary transformations as being rotations of 
a complex vector space. 

One important consequence of the fact that quantum transformations 
are unitary is that they are reversible. Thus quantum gates must be
reversible. Bennett, Fredkin, and Toffoli had already looked at
reversible versions of standard computing models 
showing that all classical computations can be done reversibly.  See Feynman's
{\em Lectures on Computation} \cite{Feynman-96} for an account of 
reversible computation and its relation to the energy of computation and information.

\subsection{Simple Quantum Gates}

The following are some examples of useful single-qubit quantum state transformations.
Because of linearity, the transformations are fully specified by
their effect on the basis vectors. 
The associated matrix,  with $\{\ket 0, \ket 1\}$ as the preferred ordered
basis, is also shown.
\begin{displaymath}
\begin{array}{ll}
\begin{array}{lrcl}
I:& \ket{0} & \to & \ket{0}\\
  & \ket{1} & \to & \ket{1}\\
\end{array} &
\left(\begin{array}{cc}1 & 0\\ 0 & 1\end{array}\right)\\
\begin{array}{lrcl}
X:& \ket{0} & \to & \ket{1}\\
  & \ket{1} & \to & \ket{0}\\
\end{array} &
\left(\begin{array}{cc}0 & 1\\ 1 & 0\end{array}\right)\\
\begin{array}{lrcl}
Y:& \ket{0} & \to &-\ket{1}\\
  & \ket{1} & \to & \ket{0}\\
\end{array} &
\left(\begin{array}{cc}0 & 1\\ -1 & 0\end{array}\right)\\
\begin{array}{lrcl}
Z:& \ket{0} & \to & \ket{0}\\
  & \ket{1} & \to & -\ket{1}\\
\end{array} &
\left(\begin{array}{cc}1 & 0\\ 0 & -1\end{array}\right)\\
\end{array}
\end{displaymath}
The names of these transformations are conventional. 
$I$ is the identity transformation, $X$ is negation, $Z$ is
a phase shift operation, and $Y = ZX$ is a combination of both.  
The $X$ transformation was discussed previously in section \ref{braket}.
It can be readily verified that these gates are unitary.  For example
\begin{displaymath}
YY^* = \left(\begin{array}{cc}0 & -1\\ 1 & 0\end{array}\right) 
	 \left(\begin{array}{cc}0 & 1\\ -1 & 0\end{array}\right) = I.
\end{displaymath}

The controlled-{\sc not}\index{controlled not} gate, $C_{not}$, 
operates on two qubits as follows: it changes the second 
bit if the first bit is $1$ and leaves this bit unchanged otherwise.
The vectors $\ket{00}$, $\ket{01}$, $\ket{10}$, and $\ket{11}$ form an orthonormal basis 
for the state space of a two-qubit system, a $4$-dimensional
complex vector space. In order to represent transformations of this
space in matrix notation we need to choose an isomorphism between 
this space and the space of complex four tuples. There is no reason,
other than convention, to pick one isomorphism over another. The one
we use here associates $\ket{00}$, $\ket{01}$, $\ket{10}$, and $\ket{11}$ to the standard
4-tuple basis $(1,0,0,0)^T$, $(0,1,0,0)^T$, $(0,0,1,0)^T$ and $(0,0,0,1)^T$,
in that order. The $C_{not}$ transformation has representations 
\begin{displaymath}
\begin{array}{ll}\begin{array}{lrcl}
C_{not}:& \ket{00} & \to & \ket{00}\\
        & \ket{01} & \to & \ket{01}\\
        & \ket{10} & \to & \ket{11}\\
        & \ket{11} & \to & \ket{10}\\
\end{array} & \left(\begin{array}{cccc}1 & 0 & 0 & 0\\ 0 & 1 & 0 & 0\\
				       0 & 0 & 0 & 1\\ 0 & 0 & 1 & 0\end{array}\right).\\
\end{array}
\end{displaymath}
The transformation $C_{not}$ is unitary since $C_{not}^*=C_{not}$ and
$C_{not}C_{not}= I$. 
The $C_{not}$ gate cannot
be decomposed into a tensor product of two 
single-bit transformations.

It is useful to have graphical representations of quantum state
transformations, especially when several transformations are
combined.
The controlled-{\sc not} gate $C_{not}$ is typically represented by a circuit of the
form
\begin{displaymath}
\begin{array}{c}\Qcontrol\\ \Rtoggle\\\end{array}.
\end{displaymath}
The open circle indicates the control bit, and the $\times$ indicates the conditional
negation of the subject bit.  In general there can be multiple control bits.  Some authors use
a solid circle to indicate negative control, in which the subject bit is toggled
when the control bit is $0$.

Similarly, the controlled-controlled-{\sc not}, which  negates the last bit
of three if and only if the first two are both $1$, has the following 
graphical representation.
\begin{displaymath}
\begin{array}{c}\Qcontrol\\ \Qcontrol\\ \Rtoggle\\\end{array}
\end{displaymath}

Single bit operations are graphically represented by 
appropriately labelled boxes as shown.

\begin{center}
\begin{picture}(10,5)(0,0)
\put(4,0){\framebox(2,2){$Z$}}
\put(1,1){\line(1,0){3}}
\put(6,1){\line(1,0){3}}
\put(4,3){\framebox(2,2){$Y$}}
\put(1,4){\line(1,0){3}}
\put(6,4){\line(1,0){3}}
\end{picture}
\end{center}

\subsubsection{The Walsh-Hadamard Transformation}

Another important single-bit transformation is the 
Hadamard\index{Hadamard} Transformation defined by
\begin{displaymath}
\begin{array}{lrcl}
H:& \ket{0} & \to & {1\over \sqrt 2}(\ket 0 + \ket 1)\\
  & \ket{1} & \to & {1\over \sqrt 2}(\ket 0 - \ket 1).\\
\end{array}
\end{displaymath}

The transformation $H$ has a number of important applications.  When 
applied to $\ket 0$, $H$ creates a superposition state 
${1\over \sqrt 2}(\ket 0 + \ket 1)$. 
\label{Walsh}
Applied to $n$ bits individually, $H$ generates a superposition of all $2^n$ 
possible states, which can be viewed as the binary representation of
the numbers from $0$ to $2^n - 1$. 
\begin{eqnarray*}
& &(H\otimes H \otimes \dots \otimes H)\ket{00\dots 0}\\
&=&{1 \over \sqrt {2^n}}\left((\ket 0+\ket 1)\otimes(\ket 0+\ket 1)\otimes\dots\otimes(\ket 0+\ket 1
)\right)\\ 
&=&{1 \over \sqrt {2^n}}\sum_{x=0}^{2^n-1}\ket x.
\end{eqnarray*}
The transformation that applies $H$ to $n$ bits is called the 
Walsh, or Walsh-Hadamard,  
transformation $W$.  It can be defined as a recursive decomposition\index{decomposition} 
of the form 
\begin{displaymath}
W_1 = H, W_{n+1} = H\otimes W_{n}.
\end{displaymath}

\subsubsection{No Cloning}
\label{NoCloning}

The unitary property implies that quantum
states cannot be copied or cloned.  The no cloning proof given here,
originally due to Wootters
and Zurek \cite{Wootters-Zurek}, is a simple application of the linearity of unitary
transformations.  

Assume that $U$ is a unitary 
transformation that clones, in that $U(\ket{a0})=\ket{aa}$ for all quantum
states $\ket a$.  Let $\ket a$ and $\ket b$ be two orthogonal quantum
states. Say $U(\ket{a0})=\ket{aa}$ and $U(\ket{b0})=\ket{bb}$.  Consider
$\ket c =(1/\sqrt 2)(\ket a+\ket b)$. By linearity, 
\begin{eqnarray*}
U(\ket{c0})&={1\over\sqrt 2}(U(\ket{a0})+U(\ket{b0}))\\
           &={1\over\sqrt 2}(\ket{aa}+\ket{bb}).
\end{eqnarray*} 
But if $U$ is a cloning transformation then
\begin{displaymath}
U(\ket{c0})=\ket{cc}=1/2(\ket{aa}+\ket{ab}+\ket{ba}+\ket{bb}),
\end{displaymath}
which is not equal to $(1/\sqrt 2)(\ket{aa}+\ket{bb})$.
Thus there is no unitary operation that can reliably clone unknown
quantum states.  It is clear that cloning is not possible by using 
measurement since measurement is both probabalistic and
destructive of states not in the measuring device's associated subspaces.

It is important to understand what sort of cloning is and isn't 
allowed. It is possible to clone a known quantum state. What the
no cloning principle tells us is that it is impossible to reliably
clone an unknown quantum state. 
Also, it is possible to obtain $n$ particles in
an entangled state $a\ket{00\dots 0}+b\ket{11\dots 1}$ from an
unknown state $a\ket{0}+b\ket{1}$. Each of these particles will behave
in exactly the same way when measured with respect to the standard
basis for quantum computation 
$\{\ket{00 \dots 0}, \ket{00 \dots 01}, \dots, \ket{1 1\dots 1}\}$, 
but not when measured with respect to other bases.
It is not possible to create the $n$ particle state
$(a\ket{0}+b\ket{1})\otimes\dots\otimes (a\ket{0}+b\ket{1})$
from an unknown state $a\ket{0}+b\ket{1}$.

\subsection{Examples}

The use of simple quantum gates can be studied with two simple examples: 
dense coding and teleportation.

\label{coding}

Dense coding\index{dense coding} uses one quantum
bit\index{quantum bit} together with an EPR pair to encode and transmit two
classical bits\index{classical bits}.  Since EPR pairs can be distributed ahead of time,
only one qubit (particle) needs to be physically transmitted to communicate two bits
of information. This result is surprising since, as was discussed in 
section \ref{information}, only one classical bit's worth of
information can be extracted from a qubit.
Teleportation\index{teleportation} is the opposite of dense
coding\index{dense coding}, in that it uses two classical bits\index{classical
bits} to transmit a single qubit.  Teleportation is surprising in light
of the no cloning principle of quantum mechanics, in that it enables
the transmission of an unknown quantum state.

The key to both dense coding\index{dense coding} and
teleportation\index{teleportation} is the use of entangled particles.
The initial set up is the same for both processes. Alice and Bob
wish to communicate. Each is sent one of the entangled particles
making up an EPR pair, 
\begin{displaymath}
\psi_0 = {1\over \sqrt 2}(\ket {00} + \ket {11}).
\end{displaymath}

Say Alice is sent the first particle, and Bob the second. So until
a particle is transmitted, only Alice can perform transformations
on her particle, and only Bob can perform transformations on his.

\subsubsection{Dense Coding}

\begin{center}
\mbox{\psfig{file=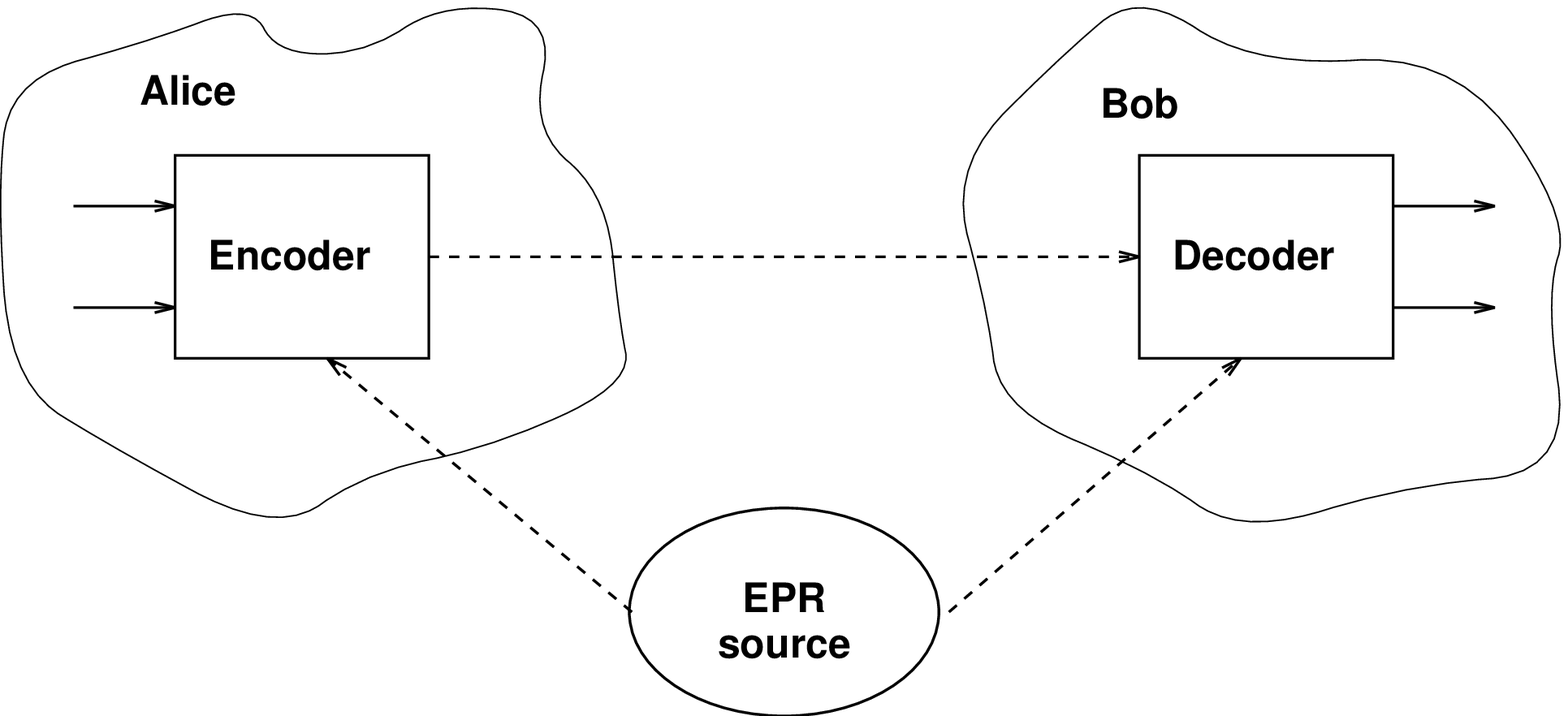,width=4in}}
\end{center}

\paragraph*{Alice}

Alice receives two classical bits\index{classical bits}, encoding the numbers $0$ through $3$.  Depending
on this number Alice performs one of the transformations $\{I, X, Y, Z\}$ on
her qubit of the entangled pair $\psi_0$.  Transforming just one bit of an
entangled pair means performing the identity transformation on the other bit.
The resulting state is shown in the table.

\begin{displaymath}
\begin{array}{ccc}
\mbox{Value} & \mbox{Transformation} & \mbox{New state} \\\hline
0 & \psi_0 = (I\otimes I) \psi_0 &  {1\over \sqrt 2}(\ket {00} + \ket {11})\\
1 & \psi_1 = (X\otimes I) \psi_0 &  {1\over \sqrt 2}(\ket {10} + \ket {01})\\
2 & \psi_2 = (Y\otimes I) \psi_0 &  {1\over \sqrt 2}(-\ket {10} + \ket {01})\\
3 & \psi_3 = (Z\otimes I) \psi_0 &  {1\over \sqrt 2}(\ket {00} - \ket {11})\\
\end{array}
\end{displaymath}

Alice then sends her qubit to Bob.

\paragraph*{Bob}

Bob applies a controlled-{\sc not}\index{controlled not} to the two qubits of the entangled pair.  

\begin{displaymath}
\begin{array}{cccc}
\mbox{Initial state} & \mbox{Controlled-{\sc not}} & \mbox{First bit} & \mbox{Second bit} \\\hline
\psi_0 = {1\over \sqrt 2}(\ket {00} + \ket {11}) &
	 {1\over \sqrt 2}(\ket {00} + \ket {10}) &
	 {1\over \sqrt 2}(\ket {0} + \ket {1}) & \ket 0\\
\psi_1 = {1\over \sqrt 2}(\ket {10} + \ket {01}) &
	 {1\over \sqrt 2}(\ket {11} + \ket {01}) &
	 {1\over \sqrt 2}(\ket {1} + \ket {0}) & \ket 1\\
\psi_2 = {1\over \sqrt 2}(-\ket {10} + \ket {01}) &
	 {1\over \sqrt 2}(-\ket {11} + \ket {01}) &
	 {1\over \sqrt 2}(-\ket {1} + \ket {0}) &  \ket 1\\
\psi_3 = {1\over \sqrt 2}(\ket {00} - \ket {11}) &
	 {1\over \sqrt 2}(\ket {00} - \ket {10}) &
	 {1\over \sqrt 2}(\ket {0} - \ket {1}) &  \ket 0\\
\end{array}
\end{displaymath}

Note that Bob can now measure the second qubit without disturbing the quantum state.  If
the measurement\index{measurement} returns $\ket 0$ then the encoded value was either $0$ or $3$, if the
measurement\index{measurement} returns $\ket 1$ then the encoded value was either $1$ or $2$.

Bob now applies $H$ to the first bit:
\begin{displaymath}
\begin{array}{ccc}
\mbox{Initial state} & \mbox{First bit} & H (\mbox{First bit}) \\\hline
\psi_0 & {1\over \sqrt 2}(\ket {0} + \ket {1}) &
	 {1\over \sqrt 2}\bigl({1\over \sqrt 2}(\ket 0 + \ket 1)
	 + {1\over \sqrt 2}(\ket 0 - \ket 1)\bigr) = \ket 0\\
\psi_1 & {1\over \sqrt 2}(\ket {1} +  \ket {0})  &
	 {1\over \sqrt 2}\bigl({1\over \sqrt 2}(\ket 0 - \ket 1)
	 + {1\over \sqrt 2}(\ket 0 + \ket 1)\bigr) = \ket 0\\
\psi_2 & {1\over \sqrt 2}(-\ket {1} + \ket {0}) &
	 {1\over \sqrt 2}\bigl(-{1\over \sqrt 2}(\ket 0 - \ket 1)
	 +{1\over \sqrt 2}(\ket 0 + \ket 1)\bigr) = \ket 1\\
\psi_3 & {1\over \sqrt 2}(\ket {0} - \ket {1}) & 
	 {1\over \sqrt 2}\bigl({1\over \sqrt 2}(\ket 0 + \ket 1)
	 - {1\over \sqrt 2}(\ket 0 - \ket 1)\bigr) = \ket 1\\
\end{array}
\end{displaymath}

Finally, Bob measures the resulting bit which allows him to distinguish between
$0$ and $3$, and $1$ and $2$.

\subsubsection{Teleportation}

The objective is to transmit the quantum state of a particle using
classical bits\index{classical bits} and reconstruct the exact quantum
state at the receiver.  Since quantum state cannot be copied, the
quantum state of the given particle will necessarily be destroyed.
Single bit teleportation\index{teleportation} has been realized
experimentally \cite{Teleportation, Nielsen, Boschi}.

\begin{center}
\mbox{\psfig{file=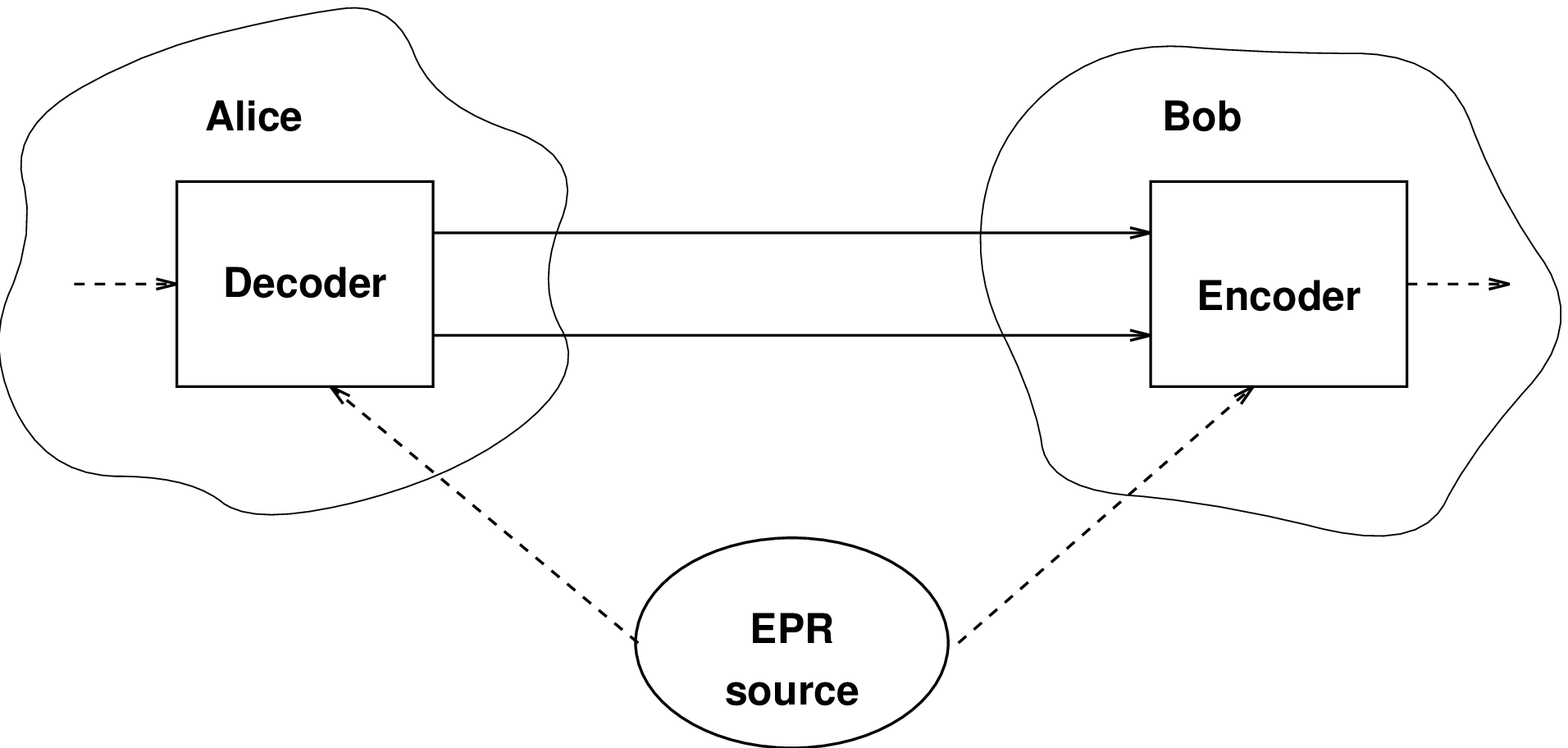,width=4in}}
\end{center}

\paragraph*{Alice}

Alice has a qubit whose state she doesn't know. She wants to send the 
state of ths qubit \begin{displaymath}
\phi = a\ket 0 + b \ket 1
					   \end{displaymath}
to Bob through classical channels.
As with dense coding, Alice and Bob each possess one qubit of an entangled pair
\begin{displaymath}
\psi_0 = {1\over \sqrt 2}(\ket {00} + \ket {11}).
\end{displaymath}

Alice applies the decoding step of dense coding to the qubit $\phi$ to be 
transmitted and her half of the entangled pair. The starting state is
quantum state 
\begin{eqnarray*}\phi\otimes \psi_0 &=&
{1\over \sqrt 2}\bigl(a \ket 0\otimes(\ket {00} + \ket {11}) + b\ket 1\otimes(\ket {00} +\ket {11})\bigr)\\
&=&{1\over \sqrt 2}\bigl(a \ket {000} +a  \ket {011} + b \ket {100} + b \ket {111}\bigr),\\
\end{eqnarray*}
of which Alice controls the first two bits and Bob 
controls the last one.  Alice now
applies $C_{not} \otimes I$ and $H \otimes I \otimes I$ to this state:
\begin{eqnarray*}
\lefteqn{(H \otimes I \otimes I)(C_{not} \otimes I)(\phi\otimes \psi_0)} \\
& = & (H \otimes I \otimes I)(C_{not} \otimes I){1\over \sqrt 2}\bigl(a \ket {000} +a  \ket {011} + b \ket {100} + b \ket {111}\bigr)\\
& = & (H \otimes I \otimes I){1\over \sqrt 2}\bigl(a\ket {000} + a\ket {011} + b\ket {110} + b\ket {101}\bigr)\\
& = & {1\over 2}\bigl(a (\ket {000} + \ket {011} +\ket {100} + \ket {111}) + b (\ket {010} + \ket {001} - \ket {110} - \ket {101})\bigr)\\
& = & {1\over 2}\bigl(\ket {00} (a \ket {0} + b \ket {1}) +
		 \ket {01} (a \ket {1} + b \ket {0}) +
		 \ket {10} (a \ket {0} - b \ket {1}) +
		 \ket {11} (a \ket {1} - b \ket {0})\bigr)\\
\end{eqnarray*}
Alice measures the first two qubits to get one of
$\ket {00}$, $\ket {01}$, $\ket {10}$, or $\ket {11}$ with equal
probability.  Depending on the result of
the measurement, the quantum state of Bob's qubit is projected to 
$a \ket {0} + b \ket {1}$, 
$a \ket {1} + b \ket {0}$,
$a \ket {0} - b \ket {1}$, or
$a \ket {1} - b \ket {0}$ respectively.  Alice sends the result 
of her measurement as two classical bits\index{classical bits} to Bob.  

Note that when she measured it, Alice irretrievably altered the state of
her original qubit $\phi$, whose state she is in the process of sending to Bob.
This loss of the original state is the reason teleportation does not
violate the no cloning principle.

\paragraph*{Bob}

When Bob receives the two classical bits\index{classical bits} from Alice 
he knows how the state of his half of the entangled pair compares 
to the original state of Alice's qubit.
\begin{displaymath}
\begin{array}{ccc}
\mbox{bits received} & \mbox{state} & \mbox{decoding} \\
00 & a \ket {0} + b \ket {1}& I\\
01 & a \ket {1} + b \ket {0}& X\\
10 & a \ket {0} - b \ket {1}& Z\\
11 & a \ket {1} - b \ket {0}& Y\\
\end{array}
\end{displaymath}

Bob can reconstruct the original state of Alice's qubit, $\phi$, by 
applying the appropriate decoding transformation to 
his part of the entangled pair.  Note that this is the encoding step
of dense coding.

\section{Quantum Computers}

\label{computer}

This section discusses how quantum mechanics can 
be used to perform computations and
how these computations are qualitatively different from those performed by a 
conventional computer.  Recall from section \ref{gates} 
that all quantum state transformations have to be
reversible.  While the classical {\sc not} gate is reversible, 
{\sc and}, {\sc or} and {\sc nand} gates are not.  Thus it is not obvious that
quantum transformations can carry out all classical computations.
The first subsection describes complete sets of reversible gates that
can perform any classical computation on a quantum computer. Furthermore,
it describes sets of gates with which all quantum computations can be
done. The second subsection discusses quantum parallelism.

\subsection{Quantum Gate Arrays}

%
%

The bra/ket notation is useful in defining complex
unitary operations. For two arbitrary 
unitary transformations $U_1$ and $U_2$, the ``conditional'' 
transformation $\ket 0\bra 0 \otimes U_1 + \ket 1\bra 1\otimes U_2$ is
also unitary.
The controlled-{\sc not}\index{controlled not} gate can defined by
\begin{displaymath}
C_{not} = \ket 0\bra 0 \otimes I + \ket 1\bra 1\otimes X.
\end{displaymath}

\label{complete-gates}
The three-bit controlled-controlled-{\sc not}\index{controlled controlled not} 
gate or Toffoli gate\index{Toffoli gate} of section \ref{gates} is also an 
instance of this conditional definition:
\begin{displaymath}
T = \ket 0\bra 0\otimes I \otimes I + \ket 1\bra 1 \otimes C_{not}.
\end{displaymath}
The Toffoli gate $T$ can be used to 
construct complete set of boolean connectives, as can
be seen from the fact that it can be used to construct the $\sc AND$ and
$\sc NOT$ operators in the
following way:
\begin{eqnarray*}
T\ket{1, 1, x} & = & \ket{1, 1, \neg x}\\
T\ket{x, y, 0} & = & \ket{x, y, x \wedge y}\\
\end{eqnarray*}
The $T$ gate is sufficient to construct arbitrary combinatorial circuits\index{combinatorial circuits}.

The following quantum circuit, for example, implements a 1 bit full adder\index{full adder}
using Toffoli and controlled-{\sc not} gates:
\begin{eqnarray*}
\ket c & \Qpass \Qcontrol \Qcontrol	\Qcontrol \Qpass \Qpass	   & \ket c\\
\ket x & \Qcontrol \Qcontrol \Qcross	\Qcross \Qcontrol \Qpass  & \ket x\\
\ket y & \Qcontrol \Qcross \Qcontrol	\Qcross \Qcross \Qcontrol  & \ket y\\
\ket 0 & \Qcross \Qcross \Qcross 	\Rtoggle \Rtoggle \Rtoggle & \ket {s}\\
\ket 0 & \Rtoggle \Rtoggle \Rtoggle	\Qpass \Qpass \Qpass       & \ket {c'}\\
\end{eqnarray*}
where $x$ and $y$ are the data bits, $s$ is their sum (modulo $2$), $c$ is the incoming carry bit, 
and $c'$ is the new carry bit.
Vedral, Barenco\index{Barenco} and Ekert\index{Ekert} \cite{Vedral-et-al-95} define more complex
circuits that include in-place addition and modular addition.

The Fredkin gate\index{Fredkin gate} is a ``controlled swap'' and can be defined as
\begin{displaymath}
F = \ket 0\bra 0\otimes I \otimes I + \ket 1\bra 1 \otimes S
\end{displaymath}
where $S$ is the swap operation
\begin{displaymath}
S = \ket{00} \bra{00} + \ket{01} \bra{10} + \ket{10} \bra{01} + \ket{11} \bra{11}.
\end{displaymath}
The reader can verify that $F$, like $T$, is complete for 
combinatorial circuits\index{combinatorial circuits}.

Deutsch has shown \cite{Deutsch-85} that it is possible to construct reversible quantum
gates for any classically computable function\index{computable function}.  
In fact, it is possible to conceive of a
universal quantum Turing machine\index{quantum Turing machine} \cite{Bernstein-Vazirani-93}.
In this construction we must assume a sufficient supply of bits that correspond to 
the tape of a Turing machine.  

Knowing that an arbitrary classical function $f$ with 
$m$ input and $k$ output bits
can be implemented on quantum computer,
we assume the existence of a {\em quantum gatearray} $U_f$ that implements $f$.
$U_f$ is a $m+k$ bit transformation of the form $U_f: \ket{x,y}\to \ket{x,y\oplus f(x)}$ where 
$\oplus$ denotes
the bitwise exclusive-{\sc or}\footnote{$\oplus$ is not the direct sum of vectors.}.
Quantum gate arrays $U_f$, defined 
in this way, are unitary \index{unitary transformation} for
any function $f$.  To compute $f(x)$ we apply $U_f$ to $\ket x$ tensored with $k$
zores $\ket{x,0}$.  Since 
$f(x)\oplus f(x) = 0$ we have $U_fU_f=I$.
Graphically the transformation $U_f: \ket {x, y} \to \ket {x, y\oplus f(x)}$ is depicted as
\begin{center}
\begin{picture}(10,10)(0,0)
\put(3,0){\framebox(4,10){$U_f$}}
\put(1,3){\line(1,0){2}}
\put(1,7){\line(1,0){2}}
\put(7,3){\line(1,0){2}}
\put(7,7){\line(1,0){2}}
\put(0.5,7){\makebox(0,0)[r]{$\ket x$}}
\put(0.5,3){\makebox(0,0)[r]{$\ket y$}}
\put(9.5,7){\makebox(0,0)[l]{$\ket x$}}
\put(9.5,3){\makebox(0,0)[l]{$\ket {y\oplus f(x)}.$}}
\end{picture}
\end{center}

While the $T$ and $F$ gates are complete for combinatorial
circuits\index{combinatorial circuits}, they cannot achieve arbitrary
quantum state transformations.  In order to realize arbitrary unitary
transformations\footnote{More precisely, we mean arbitrary unitary transformations up to a 
constant phase factor.  A constant phase shift of the state has no physical,
and therefore no computational, significance.}, 
single bit rotations need to be included.
Barenco\index{Barenco} et.~al. \cite{Barenco-et-al-95a} show that
$C_{not}$ together with all 1-bit quantum gates is a universal gate set.
It suffices to include the following one-bit transformations 
\begin{displaymath}
\left(\begin{array}{cc}\cos \alpha & \sin \alpha \\ -\sin \alpha &
\cos \alpha \end{array}\right),
\left(\begin{array}{cc}e^{i\alpha} & 0 \\ 0 & e^{-i\alpha} \end{array}\right)
\end{displaymath}
for all $0\leq\alpha\leq 2\pi$ together with the $C_{not}$ to obtain a universal set
of gates.
As we shall see, such non-classical transformations
are crucial for exploiting
the power of quantum computers.

\subsection{Quantum Parallelism}
\label{QP}

What happens if $U_f$ is applied to input which is in a 
superposition\index{superposition}?  
The answer is easy but powerful: since $U_f$ is a linear transformation,
 it is applied to all basis vectors in
the superposition simultaneously and will generate a superposition of the 
results. In this way,
it is possible to compute $f(x)$ for $n$ values of $x$ in a single
application of $U_f$. This effect is called quantum parallelism\index{quantum parallelism}.

\label{parallelism}
The power of quantum algorithms comes from taking advantage 
of quantum parallelism and entanglement. So most quantum algorithms begin by computing
a function of interest on a superposition of all values as follows.
Start with an
$n$-qubit state $\ket{00\dots 0}$. Apply the Walsh-Hadamard transformation $W$
of section \ref{Walsh}
to get a superposition
   \begin{displaymath}
\frac{1}{\sqrt{2^n}}(\ket{00 \dots 0}+\ket{00 \dots 1}+\dots +\ket{11 \dots
 1}) = \frac{1}{\sqrt{2^n}}\sum_{x=0}^{2^n-1}\ket {x}
   \end{displaymath}
which should be viewed as the superposition of all integers $0 \leq x < 2^n$.  Add
a $k$-bit register $\ket 0$ then
by linearity 
\begin{eqnarray*}
U_f(\frac{1}{\sqrt{2^n}}\sum_{x=0}^{2^n-1}\ket {x,0}) &=&
\frac{1}{\sqrt{2^n}}\sum_{x=0}^{2^n-1}U_f(\ket {x,0})\\
&=&\frac{1}{\sqrt{2^n}}\sum_{x=0}^{2^n-1}\ket {x,f(x)}\\
\end{eqnarray*}
where $f(x)$ is the function of interest. Note that since $n$ qubits
enable working simultaneously with $2^n$ states, quantum parallelism
circumvents the time/space trade-off of classical parallelism through
its ability to provide an exponential amount of computational space
in a linear amount of physical space.

Consider the trivial example of a controlled-controlled-{\sc not}\index{controlled controlled not}
\index{Toffoli gate} (Toffoli) gate, $T$, that computes
the conjunction of two values:
{\samepage
\begin{eqnarray*}
\ket x & \Qcontrol & \ket x\\
\ket y & \Qcontrol & \ket y\\
\ket 0 & \Rtoggle & \ket {x \wedge y}\\
\end{eqnarray*}
}

Now take as input a superposition\index{superposition} of all possible
bit combinations of $x$ and $y$ together with the necessary $0$:
\begin{eqnarray*}
H\ket 0\otimes H\ket 0\otimes \ket 0
 &=& {1 \over \sqrt 2}(\ket {0} + \ket {1})\otimes{1 \over \sqrt 2}(\ket {0} + \ket {1}) \otimes\ket 0\\
 &=& {1 \over 2}(\ket {000} + \ket {010} + \ket {100} + \ket {110}).\\
\end{eqnarray*}
Apply $T$ to the superposition of inputs to get a superposition of 
the results, namely
\begin{displaymath}
T(H\ket 0\otimes H\ket 0\otimes \ket 0)
= {1 \over 2}(\ket {000} + \ket {010} + \ket {100} + \ket {111}). 
\end{displaymath}
The resulting superposition\index{superposition} can be viewed as a truth table for the conjunction, or
more generally as the graph of a function. 
In the output the values of $x$, $y$, and $x \wedge y$ are entangled\index{entangled} in such a
way that measuring the result will give one line of the truth table, or more
generally one point of graph of the function.  Note that the bits can be 
measured in any order: measuring the result will project the state to a 
superposition\index{superposition} of the set of all 
input values for which $f$ produces this result and 
measuring the input will project the result to the corresponding function value.

Measuring at this point gives no advantage over classical parallelism as
only one result is obtained, and worse still one cannot even choice
which result one gets.
The heart of any quantum algorithm is the way in which it manipulates
quantum parallelism so that desired results will be measured with
high probability. This sort of manipulation has no classical analog, and
requires non-traditional programming techniques. 
We list a couple of the techniques currently known.
\begin{itemize}
\item Amplify output values of interest.  The general idea is to transform
the state in such a way that values of interest have a larger amplitude\index{amplitude} and
therefore have a higher probability of being measured. Examples of
this approach will be described in section \ref{search}.
\item Find common properties of all the values of $f(x)$.  This 
idea is exploited in Shor's algorithm\index{Shor's algorithm} 
which uses a quantum Fourier transformation\index{Fourier transformation} 
to obtain the period of $f$.
\end{itemize}

\section{Shor's Algorithm}

\label{shor}

In 1994, inspired by work of Daniel Simon (later published in \cite{Simon-94}),
Peter Shor\index{Shor} found a bounded probability 
polynomial time algorithm for factoring 
$n$-digit numbers on a quantum computer\index{quantum computer}.
Since the 1970's people have searched for efficient algorithms for 
factoring integers. The most efficient classical algorithm known 
today is that of Lenstra\index{Lenstra} and Lenstra 
\cite{Lenstra-Lenstra-93} which is exponential in
the size of the input. The input is the list of
digits of $M$, which has size $n \sim\log M$. People were confident enough
that no efficient algorithm existed, that the 
security of cryptographic\index{cryptography}
systems, like the widely used RSA algorithm, depend on the difficulty of 
this problem. Shor's result surprised the community at large,
prompting widespread interest in quantum computing.

Most factoring algorithms, including Shor's, use a standard
reduction of the factoring problem to the problem of finding the
period of a function. Shor uses quantum parallelism in the
standard way to obtain a 
superposition of all the values of the function in one step.
He then computes the quantum Fourier transform of the function,
which like classical Fourier transforms, puts all the amplitude
of the function into multiples of the reciprocal of the period.
With high probability, measuring the state 
yields the period, which in turn is used to factor the integer $M$.

The above description captures the essence of the quantum algorithm, but
is something of an oversimplification.
The biggest complication is that the quantum Fourier transform
is based on the fast Fourier transform and thus gives only approximate
results in most cases. Thus extracting the period is trickier than 
outlined above, but the techniques for extracting the period are 
classical. 

We will first describe the quantum Fourier transform and then give a 
detailed outline of Shor's algorithm.

\subsection{The Quantum Fourier Transform}

Fourier transforms in general map from the time domain to the
frequency domain.  So Fourier transforms map functions of period $r$
to functions which have non-zero values only at multiples of the
frequency $\frac{2\pi}{r}$.  Discrete Fourier transform (DFT) operates
on $N$ equally spaced samples in the interval $\lbrack 0,2\pi)$ for
some $N$ and outputs a function whose domain is
the integers between $0$ and $N-1$. 
The discrete Fourier transform of a (sampled) function of period $r$
is a function concentrated near multiples of 
$\frac{N}{r}$.  If the period $r$ divides $N$ evenly, the 
result is a function that has non-zero values only at multiples
of $\frac{N}{r}$. Otherwise, the result will approximate this behavior, 
and there will be non-zero terms 
at integers close to multiples of $\frac{N}{r}$.

The Fast Fourier transform (FFT) is a version of DFT where $N$ is a power of 2.
The quantum Fourier transform (QFT) is a variant of the discrete Fourier
transform which, like FFT, uses powers of 2.
The quantum Fourier transform operates on the amplitude of the 
quantum state, by sending
\begin{displaymath}
\sum_{x}g(x)\ket{x} \to \sum_{c}G(c)\ket{c}
\end{displaymath}
where $G(c)$ is
the discrete Fourier transform of $g(x)$, and $x$ and $c$ both range
over the binary representations for the integers between $0$ and $N-1$. 
If the state were measured after 
the Fourier transform was performed, the probability that the 
result was $\ket{c}$ would be $|G(c)|^2$.
Note that the quantum Fourier transform does not output a function 
the way the $U_f$ transformation does; no output
appears in an extra register.

Applying the quantum Fourier transform to a
periodic function $g(x)$ with period $r$, we would expect
to end up with $\sum_{c}G(c)\ket{c}$,
where $G(c)$ is zero except at multiples of $\frac{N}{r}$.
Thus, when the state is measured, the result would be a multiple
of $\frac{N}{r}$, say $j\frac{N}{r}$.
But as described above, the quantum Fourier transform
only gives approximate results for periods which are not a power of two, i.e.~do
not divide $N$.
However the larger the power of two used as a base for the transform, the
better the approximation\index{approximation}.
The quantum Fourier transform $U_{QFT}$ with base $N=2^m$ is defined by
\begin{displaymath}
U_{QFT}: \ket{x} \to \frac{1}{\sqrt{2^m}}\sum_{c=0}^{2^m-1}e^{\frac{2\pi icx}
{2^m}}\ket{c}.
\end{displaymath}

In order for Shor's algorithm to be a polynomial algorithm, the quantum
Fourier transform must be efficiently computable. Shor\index{Shor}
shows that the quantum Fourier transform with base $2^m$ 
can be constructed using
only $m(m+1)\over 2$ gates. The construction makes use of two types of
gates. One is a gate to perform the familiar Hadamard transformation $H$. 
We will denote by $H_j$ the Hadamard transformation applied to the $j$th
bit. The other type of gate performs two-bit transformations of the form
\begin{displaymath}
S_{j,k} = \left(\begin{array}{cccc}1&0&0&0\\0&1&0&0\\0&0&1&0\\0&0&0&e^{i\theta_{k-j}}\end{array}\right)
\end{displaymath}
where $\theta_{k-j}={\pi}/{2^{k-j}}$.  This transformation acts on the $k$th 
and $j$th bits of a larger register. The quantum Fourier transform
is given by 
\begin{displaymath}
H_0S_{0,1}\dots S_{0,m-1}H_1\dots H_{m-3}S_{m-3,m-2}S_{m-3,m-1}H_{m-2}S_{m-2, m-1}H_{m-1}
\end{displaymath}
followed by a bit reversal transformation.  If FFT is followed by measurement, as in 
Shor's algorithm, the bit reversal can be performed classically.  See \cite{Shor-95}
for more details.

\subsection{A Detailed Outline of Shor's algorithm}

The detailed steps of Shor's algorithm are illustrated with a running example
where we factor $M = 21$. 

\begin{description}
\item[Step 1. Quantum parallelism]
Choose an integer $a$ arbitrarily. If $a$ is not relatively
prime to $M$, we have found a factor of $M$. Otherwise apply the
rest of the algorithm.  

Let $m$ be such that $M^2 \leq 2^m < 2M^2$. [This
choice is made so that the approximation used in Step 3 for functions
whose period is not a power of $2$ 
will be good enough for the rest of the algorithm to work.]
Use quantum parallelism as described in \ref{parallelism} to
compute $f(x)=a^x \mod M$ for all integers from $0$ to $2^m-1$.  
The function is thus encoded in the quantum state
   \begin{equation}
{1\over \sqrt{2^m}}\sum_{x=0}^{2^m-1}\ket {x, f(x)}. \label{shor1}
   \end{equation}
 
\item[{Example}]  Suppose $a=11$ were randomly chosen.  Since $M^2=441 \leq 2^9 < 882 = 2M^2$
we find $m=9$.  Thus, a total of $14$ quantum bits, $9$ for $x$ and $5$ for $f(x)$
are required to compute the superposition of equation \ref{shor1}.

\item[Step 2. A state whose amplitude has the same period as $f$]
The quantum Fourier transform acts on the amplitude function associated
with the input state. In order to use the quantum Fourier transform to
obtain the period of $f$, a state is constructed whose amplitude
function has the same period as $f$. 

To construct such a state,
measure the last $\lceil log_2 M\rceil$ qubits of the state of equation \ref{shor1} that encode 
$f(x)$. A random value $u$ is obtained. The
value $u$ is not of interest in itself; only the
effect the measurement has on our set of superpositions is of interest. 
This measurement projects the state space onto the
subspace compatible with the measured value, so the state after
measurement is 
\begin{displaymath}
C\sum_{x}g(x)\ket{x,u},
\end{displaymath}
for some scale factor $C$ where 
\begin{displaymath}
g(x) = \left\{ \begin{array}{ll}
                1  & \mbox{if $f(x)=u$} \\
                0  & \mbox{otherwise.}
                \end{array}
        \right.
\end{displaymath}
Note that the $x$'s that actually appear in the sum, those with $g(x)\ne 0$,
differ from each other by multiples of the period, thus $g(x)$ is the
function we are looking for. If we could measure
two successive $x$'s in the sum, we would have the period. 
Unfortunately the laws of quantum physics permit only one measurement.

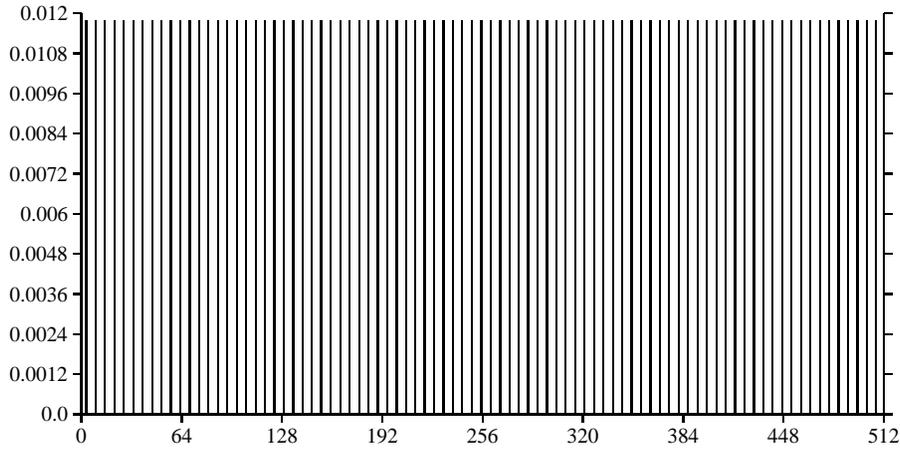
\begin{figure}
\begin{center}
\setlength{\unitlength}{0.7in}
\begin{picture}(6.600000000000001,3.3)(-0.6000000000000001,-0.3)
\put(0,0){\line(0,1){3}}
\put(-0.1,0.0){\makebox(0,0)[cr]{0.0}}\put(-0.06,0.0){\line(1,0){0.06}}
\put(-0.1,0.3){\makebox(0,0)[cr]{0.0012}}\put(-0.06,0.3){\line(1,0){0.06}}
\put(-0.1,0.6){\makebox(0,0)[cr]{0.0024}}\put(-0.06,0.6){\line(1,0){0.06}}
\put(-0.1,0.9){\makebox(0,0)[cr]{0.0036}}\put(-0.06,0.9){\line(1,0){0.06}}
\put(-0.1,1.2){\makebox(0,0)[cr]{0.0048}}\put(-0.06,1.2){\line(1,0){0.06}}
\put(-0.1,1.5){\makebox(0,0)[cr]{0.006}}\put(-0.06,1.5){\line(1,0){0.06}}
\put(-0.1,1.8){\makebox(0,0)[cr]{0.0072}}\put(-0.06,1.8){\line(1,0){0.06}}
\put(-0.1,2.1){\makebox(0,0)[cr]{0.0084}}\put(-0.06,2.1){\line(1,0){0.06}}
\put(-0.1,2.4){\makebox(0,0)[cr]{0.0096}}\put(-0.06,2.4){\line(1,0){0.06}}
\put(-0.1,2.7){\makebox(0,0)[cr]{0.0108}}\put(-0.06,2.7){\line(1,0){0.06}}
\put(-0.1,3.0){\makebox(0,0)[cr]{0.012}}\put(-0.06,3.0){\line(1,0){0.06}}
\put(6,0){\line(0,1){3}}
\put(6.0,0.0){\line(1,0){0.06}}
\put(6.0,0.3){\line(1,0){0.06}}
\put(6.0,0.6){\line(1,0){0.06}}
\put(6.0,0.9){\line(1,0){0.06}}
\put(6.0,1.2){\line(1,0){0.06}}
\put(6.0,1.5){\line(1,0){0.06}}
\put(6.0,1.8){\line(1,0){0.06}}
\put(6.0,2.1){\line(1,0){0.06}}
\put(6.0,2.4){\line(1,0){0.06}}
\put(6.0,2.7){\line(1,0){0.06}}
\put(6.0,3.0){\line(1,0){0.06}}
\put(0,0){\line(1,0){6}}
\put(0.0,-0.1){\makebox(0,0)[tc]{0}}\put(0.0,-0.06){\line(0,1){0.06}}
\put(0.75,-0.1){\makebox(0,0)[tc]{64}}\put(0.75,-0.06){\line(0,1){0.06}}
\put(1.5,-0.1){\makebox(0,0)[tc]{128}}\put(1.5,-0.06){\line(0,1){0.06}}
\put(2.25,-0.1){\makebox(0,0)[tc]{192}}\put(2.25,-0.06){\line(0,1){0.06}}
\put(3.0,-0.1){\makebox(0,0)[tc]{256}}\put(3.0,-0.06){\line(0,1){0.06}}
\put(3.75,-0.1){\makebox(0,0)[tc]{320}}\put(3.75,-0.06){\line(0,1){0.06}}
\put(4.5,-0.1){\makebox(0,0)[tc]{384}}\put(4.5,-0.06){\line(0,1){0.06}}
\put(5.25,-0.1){\makebox(0,0)[tc]{448}}\put(5.25,-0.06){\line(0,1){0.06}}
\put(6.0,-0.1){\makebox(0,0)[tc]{512}}\put(6.0,-0.06){\line(0,1){0.06}}
\put( 0.035,0){\line(0,1){2.9412}}
\put( 0.105,0){\line(0,1){2.9412}}
\put( 0.176,0){\line(0,1){2.9412}}
\put( 0.246,0){\line(0,1){2.9412}}
\put( 0.316,0){\line(0,1){2.9412}}
\put( 0.387,0){\line(0,1){2.9412}}
\put( 0.457,0){\line(0,1){2.9412}}
\put( 0.527,0){\line(0,1){2.9412}}
\put( 0.598,0){\line(0,1){2.9412}}
\put( 0.668,0){\line(0,1){2.9412}}
\put( 0.738,0){\line(0,1){2.9412}}
\put( 0.809,0){\line(0,1){2.9412}}
\put( 0.879,0){\line(0,1){2.9412}}
\put( 0.949,0){\line(0,1){2.9412}}
\put(1.0195,0){\line(0,1){2.9412}}
\put(1.0898,0){\line(0,1){2.9412}}
\put(1.1602,0){\line(0,1){2.9412}}
\put(1.2305,0){\line(0,1){2.9412}}
\put(1.3008,0){\line(0,1){2.9412}}
\put(1.3711,0){\line(0,1){2.9412}}
\put(1.4414,0){\line(0,1){2.9412}}
\put(1.5117,0){\line(0,1){2.9412}}
\put( 1.582,0){\line(0,1){2.9412}}
\put(1.6523,0){\line(0,1){2.9412}}
\put(1.7227,0){\line(0,1){2.9412}}
\put( 1.793,0){\line(0,1){2.9412}}
\put(1.8633,0){\line(0,1){2.9412}}
\put(1.9336,0){\line(0,1){2.9412}}
\put(2.0039,0){\line(0,1){2.9412}}
\put(2.0742,0){\line(0,1){2.9412}}
\put(2.1445,0){\line(0,1){2.9412}}
\put(2.2148,0){\line(0,1){2.9412}}
\put(2.2852,0){\line(0,1){2.9412}}
\put(2.3555,0){\line(0,1){2.9412}}
\put(2.4258,0){\line(0,1){2.9412}}
\put(2.4961,0){\line(0,1){2.9412}}
\put(2.5664,0){\line(0,1){2.9412}}
\put(2.6367,0){\line(0,1){2.9412}}
\put( 2.707,0){\line(0,1){2.9412}}
\put(2.7773,0){\line(0,1){2.9412}}
\put(2.8477,0){\line(0,1){2.9412}}
\put( 2.918,0){\line(0,1){2.9412}}
\put(2.9883,0){\line(0,1){2.9412}}
\put(3.0586,0){\line(0,1){2.9412}}
\put(3.1289,0){\line(0,1){2.9412}}
\put(3.1992,0){\line(0,1){2.9412}}
\put(3.2695,0){\line(0,1){2.9412}}
\put(3.3398,0){\line(0,1){2.9412}}
\put(3.4102,0){\line(0,1){2.9412}}
\put(3.4805,0){\line(0,1){2.9412}}
\put(3.5508,0){\line(0,1){2.9412}}
\put(3.6211,0){\line(0,1){2.9412}}
\put(3.6914,0){\line(0,1){2.9412}}
\put(3.7617,0){\line(0,1){2.9412}}
\put( 3.832,0){\line(0,1){2.9412}}
\put(3.9023,0){\line(0,1){2.9412}}
\put(3.9727,0){\line(0,1){2.9412}}
\put( 4.043,0){\line(0,1){2.9412}}
\put(4.1133,0){\line(0,1){2.9412}}
\put(4.1836,0){\line(0,1){2.9412}}
\put(4.2539,0){\line(0,1){2.9412}}
\put(4.3242,0){\line(0,1){2.9412}}
\put(4.3945,0){\line(0,1){2.9412}}
\put(4.4648,0){\line(0,1){2.9412}}
\put(4.5352,0){\line(0,1){2.9412}}
\put(4.6055,0){\line(0,1){2.9412}}
\put(4.6758,0){\line(0,1){2.9412}}
\put(4.7461,0){\line(0,1){2.9412}}
\put(4.8164,0){\line(0,1){2.9412}}
\put(4.8867,0){\line(0,1){2.9412}}
\put( 4.957,0){\line(0,1){2.9412}}
\put(5.0273,0){\line(0,1){2.9412}}
\put(5.0977,0){\line(0,1){2.9412}}
\put( 5.168,0){\line(0,1){2.9412}}
\put(5.2383,0){\line(0,1){2.9412}}
\put(5.3086,0){\line(0,1){2.9412}}
\put(5.3789,0){\line(0,1){2.9412}}
\put(5.4492,0){\line(0,1){2.9412}}
\put(5.5195,0){\line(0,1){2.9412}}
\put(5.5898,0){\line(0,1){2.9412}}
\put(5.6602,0){\line(0,1){2.9412}}
\put(5.7305,0){\line(0,1){2.9412}}
\put(5.8008,0){\line(0,1){2.9412}}
\put(5.8711,0){\line(0,1){2.9412}}
\put(5.9414,0){\line(0,1){2.9412}}
\end{picture}
\caption{Probabilities for measuring $x$ when measuring the state 
$C\sum_{x\in X}\ket{x,8}$ obtained in Step 2, where 
$X = \{x|2 11^x \mbox{ mod } 21 = 8\}\}$}
\label{shorfig1}
\end{center}
\end{figure}

\item[{Example}] Suppose that random measurement of the superposition of equation \ref{shor1}
produces $8$.  The state after this measurement\footnote{Only the $9$ bits of $x$ 
are shown in Figure \ref{shorfig1}; the bits of $f(x)$ are known from the measurement.}
(Figure \ref{shorfig1}) clearly shows the periodicity of $f$.

\item[Step 3. Applying a quantum Fourier transform]

The $\ket u$ part
of the state will not be used, so we will no longer write it.
Apply the quantum Fourier transform to the state obtained in Step 2.
\begin{displaymath}
U_{QFT}:\sum_{x}g(x)\ket{x} \to \sum_{c}G(c)\ket{c}
\end{displaymath}

Standard Fourier analysis tells us that when the
period $r$ of the function $g(x)$ defined in Step 2 is a power of 
two, the result of the quantum Fourier
transform is \begin{displaymath}
\sum_{j}c_j\ket{j\frac{2^m}{ r}},
	     \end{displaymath}
where the amplitude is $0$ except at multiples of $2^m/r$.
When the period $r$ does not divide $2^m$, the transform approximates
the exact case so most of the amplitude
is attached to integers close to multiples of $\frac{2^m}{r}$.

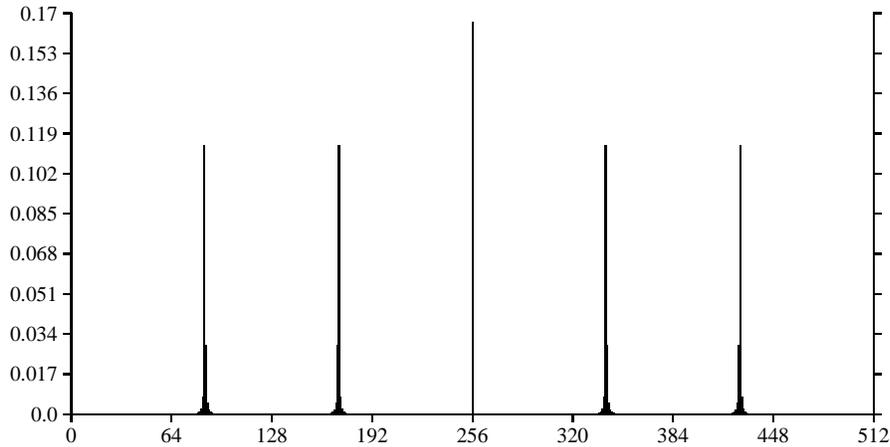
\begin{figure}
\begin{center}
\setlength{\unitlength}{0.7in}
\begin{picture}(6.600000000000001,3.3)(-0.6000000000000001,-0.3)
\put(0,0){\line(0,1){3}}
\put(-0.1,0.0){\makebox(0,0)[cr]{0.0}}\put(-0.06,0.0){\line(1,0){0.06}}
\put(-0.1,0.3){\makebox(0,0)[cr]{0.017}}\put(-0.06,0.3){\line(1,0){0.06}}
\put(-0.1,0.6){\makebox(0,0)[cr]{0.034}}\put(-0.06,0.6){\line(1,0){0.06}}
\put(-0.1,0.9){\makebox(0,0)[cr]{0.051}}\put(-0.06,0.9){\line(1,0){0.06}}
\put(-0.1,1.2){\makebox(0,0)[cr]{0.068}}\put(-0.06,1.2){\line(1,0){0.06}}
\put(-0.1,1.5){\makebox(0,0)[cr]{0.085}}\put(-0.06,1.5){\line(1,0){0.06}}
\put(-0.1,1.8){\makebox(0,0)[cr]{0.102}}\put(-0.06,1.8){\line(1,0){0.06}}
\put(-0.1,2.1){\makebox(0,0)[cr]{0.119}}\put(-0.06,2.1){\line(1,0){0.06}}
\put(-0.1,2.4){\makebox(0,0)[cr]{0.136}}\put(-0.06,2.4){\line(1,0){0.06}}
\put(-0.1,2.7){\makebox(0,0)[cr]{0.153}}\put(-0.06,2.7){\line(1,0){0.06}}
\put(-0.1,3.0){\makebox(0,0)[cr]{0.17}}\put(-0.06,3.0){\line(1,0){0.06}}
\put(6,0){\line(0,1){3}}
\put(6.0,0.0){\line(1,0){0.06}}
\put(6.0,0.3){\line(1,0){0.06}}
\put(6.0,0.6){\line(1,0){0.06}}
\put(6.0,0.9){\line(1,0){0.06}}
\put(6.0,1.2){\line(1,0){0.06}}
\put(6.0,1.5){\line(1,0){0.06}}
\put(6.0,1.8){\line(1,0){0.06}}
\put(6.0,2.1){\line(1,0){0.06}}
\put(6.0,2.4){\line(1,0){0.06}}
\put(6.0,2.7){\line(1,0){0.06}}
\put(6.0,3.0){\line(1,0){0.06}}
\put(0,0){\line(1,0){6}}
\put(0.0,-0.1){\makebox(0,0)[tc]{0}}\put(0.0,-0.06){\line(0,1){0.06}}
\put(0.75,-0.1){\makebox(0,0)[tc]{64}}\put(0.75,-0.06){\line(0,1){0.06}}
\put(1.5,-0.1){\makebox(0,0)[tc]{128}}\put(1.5,-0.06){\line(0,1){0.06}}
\put(2.25,-0.1){\makebox(0,0)[tc]{192}}\put(2.25,-0.06){\line(0,1){0.06}}
\put(3.0,-0.1){\makebox(0,0)[tc]{256}}\put(3.0,-0.06){\line(0,1){0.06}}
\put(3.75,-0.1){\makebox(0,0)[tc]{320}}\put(3.75,-0.06){\line(0,1){0.06}}
\put(4.5,-0.1){\makebox(0,0)[tc]{384}}\put(4.5,-0.06){\line(0,1){0.06}}
\put(5.25,-0.1){\makebox(0,0)[tc]{448}}\put(5.25,-0.06){\line(0,1){0.06}}
\put(6.0,-0.1){\makebox(0,0)[tc]{512}}\put(6.0,-0.06){\line(0,1){0.06}}
\put(   0.0,0){\line(0,1){2.9297}}
\put( 0.949,0){\line(0,1){ 0.011}}
\put( 0.961,0){\line(0,1){ 0.019}}
\put( 0.973,0){\line(0,1){  0.04}}
\put( 0.984,0){\line(0,1){ 0.124}}
\put( 0.996,0){\line(0,1){  2.01}}
\put(1.0078,0){\line(0,1){  0.51}}
\put(1.0195,0){\line(0,1){ 0.083}}
\put(1.0313,0){\line(0,1){ 0.033}}
\put( 1.043,0){\line(0,1){ 0.018}}
\put(1.0547,0){\line(0,1){ 0.011}}
\put(1.9453,0){\line(0,1){ 0.011}}
\put( 1.957,0){\line(0,1){ 0.018}}
\put(1.9688,0){\line(0,1){ 0.033}}
\put(1.9805,0){\line(0,1){ 0.083}}
\put(1.9922,0){\line(0,1){  0.51}}
\put(2.0039,0){\line(0,1){  2.01}}
\put(2.0156,0){\line(0,1){ 0.124}}
\put(2.0273,0){\line(0,1){  0.04}}
\put(2.0391,0){\line(0,1){ 0.019}}
\put(2.0508,0){\line(0,1){ 0.011}}
\put(   3.0,0){\line(0,1){2.9297}}
\put(3.9492,0){\line(0,1){ 0.011}}
\put(3.9609,0){\line(0,1){ 0.019}}
\put(3.9727,0){\line(0,1){  0.04}}
\put(3.9844,0){\line(0,1){ 0.124}}
\put(3.9961,0){\line(0,1){  2.01}}
\put(4.0078,0){\line(0,1){  0.51}}
\put(4.0195,0){\line(0,1){ 0.083}}
\put(4.0313,0){\line(0,1){ 0.033}}
\put( 4.043,0){\line(0,1){ 0.018}}
\put(4.0547,0){\line(0,1){ 0.011}}
\put(4.9453,0){\line(0,1){ 0.011}}
\put( 4.957,0){\line(0,1){ 0.018}}
\put(4.9688,0){\line(0,1){ 0.033}}
\put(4.9805,0){\line(0,1){ 0.083}}
\put(4.9922,0){\line(0,1){  0.51}}
\put(5.0039,0){\line(0,1){  2.01}}
\put(5.0156,0){\line(0,1){ 0.124}}
\put(5.0273,0){\line(0,1){  0.04}}
\put(5.0391,0){\line(0,1){ 0.019}}
\put(5.0508,0){\line(0,1){ 0.011}}
\end{picture}
\caption{Probability distribution of the quantum state after Fourier Transformation.}
\label{shorfig2}
\end{center}
\end{figure}

\item[{Example}] Figure \ref{shorfig2} shows the result of applying the quantum
Fourier Transform to the state obtained in Step 2. Note that 
Figure \ref{shorfig2} is the graph of the fast Fourier transform of
the function shown in  Figure \ref{shorfig1}.  In this particular
example the period of $f$ does not divide $2^m$.  

\item[Step 4. Extracting the period]

Measure the state in the standard basis for quantum computation, 
and call the result $v$.  In the case 
where the period happens to be a power of $2$, so that
the quantum Fourier transform gives exactly multiples of $2^m/r$, 
the period is easy to extract. In this case,
$v=j\frac{2^m}{r}$ for some $j$. Most of the time $j$ and $r$ will
be relatively prime, in which case reducing the fraction $\frac{v}{2^m} (= \frac jr)$
to its lowest terms will yield a fraction whose denominator $q$
is the period $r$. The fact that in general the quantum Fourier
transform only approximately gives multiples of the scaled
frequency complicates the extraction of the period from the
measurement. When the period is not a power of $2$, a good guess for
the period can be obtained using the continued fraction expansion of
$\frac{v}{2^m}$. This classical technique is described in Appendix
\ref{continued fractions}.

\item[{Example}] Say that measurement of 
the state returns $v = 427$.  Since $v$ and 
$2^m$ are relative prime the period $r$ will most likely not divide $2^m$ and
the continued fraction expansion described in Appendix \ref{continued fractions}
needs to be applied.  The following is a trace of the algorithm described
in Appendix \ref{continued fractions}:
\[
\begin{array}{r|c|c|c|c}
i & a_i & p_i & q_i & \epsilon_i \\\hline
0 &   0 &   0 &   1 & 0.8339844\\
1 &   1 &   1 &   1 & 0.1990632\\
2 &   5 &   5 &   6 & 0.02352941\\
3 &  42 & 211 & 253 &  0.5\\
\end{array}
\]
which terminates with $6 = q_2 < M \leq q_3$.  Thus, $q = 6$ is likely to be the period of $f$. 

\item[Step 5. Finding a factor of $M$]
When our guess for the period, $q$, is even, use the 
Euclidean algorithm\index{Euclidean algorithm} 
to efficiently check whether either $a^{q/2}+1$ or $a^{q/2}-1$ has 
a non-trivial common factor with $M$.

The reason why  $a^{q/2}+1$ or $a^{q/2}-1$ is likely to have a non-trivial
common factor with $M$ is as follows. If $q$ is indeed the period
of $f(x)=a^x\mod M$, then $a^q = 1\mod M$ since $a^qa^x=a^x\mod M$ for all $x$. 
If $q$ is even, we can write
\begin{displaymath}
(a^{q/2}+1)(a^{q/2}-1)=0\mod M.
\end{displaymath}
Thus, so long as neither $a^{q/2}+1$ nor $a^{q/2}-1$ is a multiple of $M$,
either $a^{q/2}+1$ or  $a^{q/2}-1$ has a non-trivial common factor with $M$.

\item[{Example}]
Since $6$ is even either $a^{6/2}-1=11^3-1 = 1330$ or $a^{6/2}+1=11^3+1=1332$ will have a common factor with $M$.  
In this particular example we find two factors $\mbox{gcd}(21, 1330) = 7$ and 
$\mbox{gcd}(21, 1332) = 3$.

\item[Step 6. Repeating the algorithm, if necessary]

Various things could have gone wrong so that this process does 
not yield a factor of $M$:
\begin{enumerate}
\item The value $v$ was not close enough to a multiple of $\frac{2^m}{r}$.
\item The period $r$ and the multiplier $j$ could have had a common factor
so that the denominator $q$ was actually a factor of the period not the 
period itself.
\item Step 5 yields $M$ as $M$'s factor.
\item The period of $f(x) = a^x \mod M$ is odd.
\end{enumerate}
Shor shows that few repetitions of this algorithm yields a factor of $M$ with
high probability.
\end{description}

\subsubsection{A Comment on Step 2 of Shor's Algorithm}
\label{eliminating2}
The measurement in Step 2 can be skipped entirely. More generally
Bernstein and Vazirani \cite{Bernstein-Vazirani-93} show that 
measurements in the middle of an algorithm can always be avoided. 
If the measurement in Step 2 is omitted, the state consists of a 
superpositions of several periodic functions all of which have the
same period.  By the linearity of quantum algorithms, applying 
the quantum Fourier transformation leads to a superposition of 
the Fourier transforms of these functions, each of which
is entangled with the corresponding $u$ and therefore do not
interfere with each other. Measurement gives a value from
one of these Fourier transforms.
Seeing how this argument can be formalized 
illustrates some of the subtleties of working with quantum 
superpostions. Apply the quantum Fourier
transform tensored with the identity, $U_{QFT}\otimes I$, 
to $C\sum_{x=0}^{2^n-1}\ket {x, f(x)}$ to get
\begin{displaymath}
C'\sum_{x=0}^{2^n-1}\sum_{c=0}^{2^m-1}e^{\frac{2\pi ixc}{2^m}}\ket{c,f(x)},
\end{displaymath}
which is equal to 
\begin{displaymath}
C'\sum_{u}\sum_{x|f(x)=u}\sum_{c}e^{\frac{2\pi ixc}{2^m}}\ket{c,u}
\end{displaymath}
for $u$ in the range of $f(x)$.
What results is a superposition of the results of Step 3 for all possible 
$u$'s.  The quantum Fourier transform is being applied to a family of separate
functions $g_u$ indexed by $u$ where
\begin{displaymath}
g_u = \left\{ \begin{array}{ll}
                1  & \mbox{if $f(x)=u$} \\
                0  & \mbox{otherwise,}
                \end{array}
        \right.
\end{displaymath}
all with the same period.
Note that the amplitudes in states with different $u$'s never
interfere (add or cancel) with each other.
The transform $U_{QFT}\otimes I$ as applied above can be written
\begin{displaymath}
U_{QFT}\otimes I:C\sum_{u\in R}\sum_{x=0}^{2^n-1}g_u(x)\ket{x,f(x)} \to C'\sum_{u\in R}\sum_{x=0}^{2^n-1}\sum_{c=0}^{2^n-1}G_u(c)\ket{c,u},
\end{displaymath}
where $G_u(c)$ is the discrete Fourier transform of $g_u(x)$ and $R$ is the
range of $f(x)$.

Measure $c$ and run Steps 4 and 5 as before.

\section{Search Problems}

\label{search}

A large class of problems can be specified as search 
problems\index{search problems} of the form
``find some $x$ in a set of possible solutions such that statement $P(x)$ is true.''
Such problems range from  database search\index{database search} 
to sorting to graph coloring. 
For example, the graph coloring problem can be viewed as a search
for an assignment of colors to vertices so that the statement ``all
adjacent vertices have different colors" is true. Similarly, a
sorting problem can be viewed as a search for a permutation for which
the statement ``the permutation $x$ takes the initial state to the
desired sorted state" is true.

An {\em unstructured} search problem is one where nothing is know (or no
assumption are used) about the structure of the solution space and the
statement $P$.  For example, determining $P(x_0)$ provides no information about
the possible value of $P(x_1)$ for $x_0\not=x_1$.  A {\em structured}
search problem is one where information about the search space and 
statement $P$ can be exploited.

For instance, searching an alphabetized list is a structured search
problem and the structure can be exploited to construct efficient
algorithms.  In other cases, like constraint satisfaction problems
such as 3-SAT or graph colorability, the problem structure can be
exploited for heuristic algorithms that yield efficient solution for
some problem instances.  But in the general case of an unstructured
problem, randomly testing the truth of statements $P (x_i)$ one by one
is the best that can be done classically.  For a search space of size
$N$, the general unstructured search problem requires $O(N)$
evaluations of $P$.  On a quantum computer, however, Grover showed
that the unstructured search problem can be solved with bounded
probability within $O(\sqrt{N})$ evaluations of $P$. Thus
Grover\index{Grover}'s search algorithm \cite{STOC::Grover1996} is
provably more efficient than any algorithm that could run on a
classical computer.

While Grover's algorithm is optimal
\cite{Bennett-et-al-97,Boyer-et-al-96,Zalka-97} for completely
unstructured searches, most search problems involve searching a
structured solution space.  Just as there are classical heuristic algorithms 
that exploit problem structure, one would expect that there are more
efficient quantum algorithms for certain structured problem instances.  
Grover et.al.~\cite{Cerf-et.al.} uses Grover's search algorithm
in place of classical searches within a heuristic algorithm to show
that a quadratic speed-up is possible over a particularly simple classical
heuristic for solving NP-hard problems.  
Brassard et.al.~\cite{Brassard-Hoyer-Tapp}, using the techniques
of Grover's search algorithm in a less obvious way, 
show that general heuristic searches have quantum analogs with quadratic 
speed-up. 

There is hope that for certain structured problems a speed-up
greater than quadratic is possible.  Such algorithms will likely
require new approaches that are not merely quantum implementations of
classical algorithms.  Shor's algorithm, when viewed as a search for
factors, is an example of an algorithm that achieves exponential
speed-up by using problem structure (number theory) in new ways unique
to quantum computation.

Tad Hogg has developed heuristic quantum search algorithms that 
exploit problem structure.  His approach is distincly non-classical and
uses unique properties of quantum computation.
One problem with this approach is that, like most heuristic algorithms,
the use of problem structure is complicated enough that it is
hard to determine the probability that a single iteration of 
an algorithm will give a correct answer. Therefore it is unknown
how efficient Hogg's algorithms are. 
Classically the efficiency of heuristic algorithms is estimated
by empirically testing the algorithm. But as there is an 
exponential slow down when simulating a quantum computer on a 
classical one, empirical testing of quantum algorithms is currently
infeasible except in small cases. 
Small cases indicate that Hogg's
algorithms are more efficient than Grover's algorithm applied to
structured search problems, but that the speed-up is likely to 
be only polynomial. While less interesting theoretically, even a 
small polynomial speed-up on average for these computational
difficult problems is of significant practical interest. Until 
sufficiently large quantum computers are
built, or better techniques for analyzing such algorithms are found,
the efficiency cannot be determined for sure.

\subsection{Grover's Search Algorithm}

Grover\index{Grover}'s algorithm 
searches an unstructured list of size $N$ for an $x$ that
makes a statement true.
Let $n$ be such that $2^n \geq N$, and let $U_p$ be the quantum 
gate that implements the classical function $P(x)$ that tests
the truth of the statement, where true is encoded as $1$.
\begin{displaymath}
U_P: \ket {x, 0} \to \ket{x, P(x)}
\end{displaymath}
The first step is the standard one for quantum computing described in
section \ref{QP}.
Compute $P$ for all possible inputs $x_i$, by
applying $U_P$ to a register containing the 
superposition ${1\over \sqrt{2^n}}\sum_{x=0}^{n-1}\ket{x}$
of all $2^n$ possible inputs $x$ together with a register set to $0$, leading
to the superposition
\begin{equation}
	{1\over \sqrt{2^n}}\sum_{x=0}^{n-1}\ket{x, P(x)}.\label{sup}
\end{equation}
The difficult step is to obtain a useful result from this superposition.

For any $x_0$ such that $P(x_0)$ is true, $\ket{x_0, 1}$ will be part of the 
superposition of Eq.~\ref{sup}. Since the  
amplitude\index{amplitude} of such a state is ${1\over \sqrt{2^n}}$, the
probability that a random measurement\index{measurement}
of the superposition produces $x_0$ is only $2^{-n}$.  
The trick is to change the quantum state in Eq.~\ref{sup} 
so as to greatly increase
the amplitude of vectors $\ket{x_0, 1}$ for which $P$ is true and 
decrease the amplitude of vectors $\ket{x, 0}$ for which $P$ is false.

Once such a transformation of the quantum state has been performed, one can 
simply measure the last qubit of the quantum state which represents $P(x)$.  
Because of the amplitude 
change, there is a high probability that the result will be $1$.  If this is the
case, the measurement\index{measurement} has projected the state of Eq.~\ref{sup}
onto the subspace ${1\over \sqrt{2^k}}\sum_{i=1}^{k}\ket{x_i, 1}$ where $k$ is the 
number of solutions.  Further measurement of the remaining bits will provide one of these solutions.
If the measurement of qubit $P(x)$ yields $0$, then the whole process is started over
and the superposition of Eq.~\ref{sup} must be computed again.

Grover\index{Grover}'s algorithm then consists of the following steps:
\begin{enumerate}
\item Prepare a register containing a superposition of all possible values $x_i\in [0\dots 2^n-1]$.
\item Compute $P(x_i)$ on this register.
\item Change amplitude $a_j$ to $-a_j$ for $x_j$ such that $P(x_j)=1$. An
efficient algorithm for changing selected signs is described in section
\ref{ChangingSigns}. A plot of the amplitudes after this step is shown 
here.

\begin{center}\mbox{\psfig{file=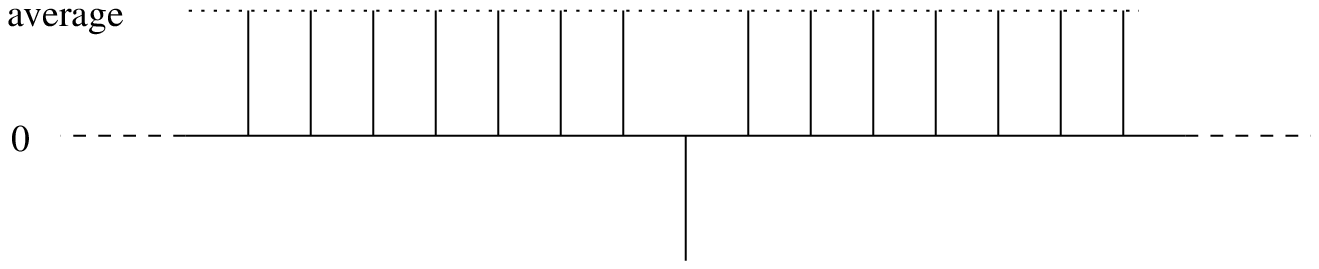,width=3in}}\end{center}

\item Apply inversion\index{inversion} about the average to increase 
amplitude of $x_j$ with $P(x_j)=1$. The quantum algorithm to
efficiently perform inversion about the average is given in
section \ref{inversion}. The resulting amplitudes look as shown, where
the amplitude of all the $x_i$'s with $P(x_i)=0$ have been diminished
imperceptibly.

\begin{center}\mbox{\psfig{file=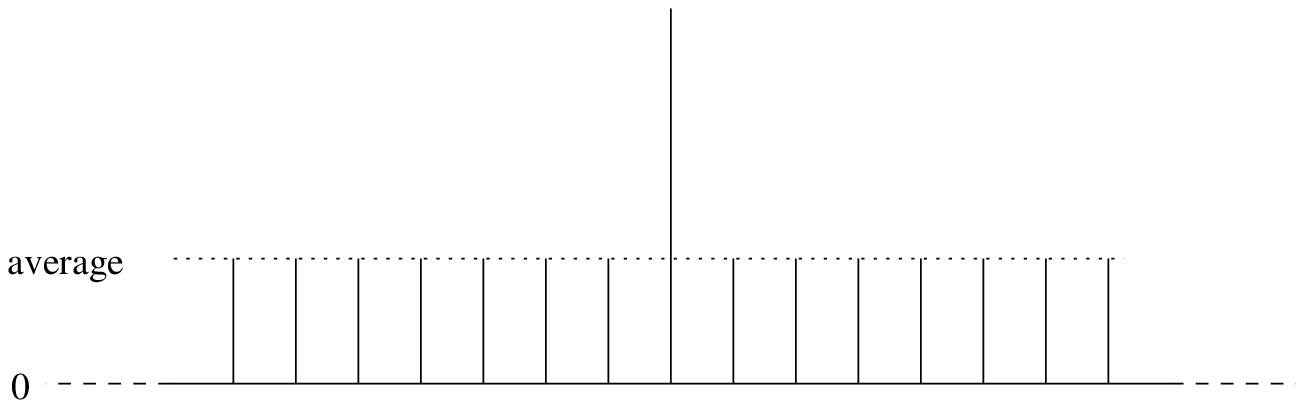,width=3in}}\end{center}

\item Repeat steps 2 through 4 ${\pi\over 4}\sqrt{2^n}$ times.
\item Read the result.
\end{enumerate}

Boyer\index{Boyer} et.al.~\cite{Boyer-et-al-96} provide 
a detailed analysis of the 
performance of Grover\index{Grover}'s algorithm.
They prove that Grover\index{Grover}'s algorithm is 
optimal up to a constant factor; no
quantum algorithm\index{quantum algorithm} can perform an unstructured 
search faster.
They also show that if there is
only a single $x_0$ such that $P(x_0)$ is true, then 
after ${\pi\over 8}\sqrt {2^n}$ iterations of steps 2 through 4 the 
failure rate is $0.5$.
After iterating ${\pi\over 4}\sqrt {2^n}$ times the failure rate drops to $2^{-n}$.  
Interestingly, additional iterations will increase the failure 
rate.  For example, after
${\pi\over 2}\sqrt {2^n}$ iterations the failure rate is close to $1$.  

There are many classical algorithms in which a procedure is repeated
over and over again for ever better results. Repeating quantum procedures
may improve results for a while, but after a sufficient number of 
repetitions the results will get worse again. Quantum procedures are
unitary transformations, which are rotations of complex space, and thus
while a repeated applications of a quantum transform may rotate the
state closer and closer to the desired state for a while, eventually
it will rotate past the desired state to get farther and farther from
the desired state. Thus to obtain useful results from a repeated 
application of a quantum transformation, one must know when to stop.
Brassard et.al.~\cite{Brassard-Hoyer-Tapp} describe an extension of Grover's algorithm
that uses Fourier Transforms to determine the number of solutions and 
the optimal number of iterations.  The extension does not increase the
overall complexity of the algorithm.

Grover has extended his algorithm to achieve quadratic speed-up for
other non-search problems such as computing the mean and median of a
function \cite{Grover-xxx}.  Using similar techniques grover has also
shown that certain search problems that classically run in $O(\log N)$
can be solved in $O(1)$ on a quantum computer.  Grover's search can used as a 
subroutine in other quantum computations since Biron et.al.~\cite{Biron-et-al} 
show how the technique can be used with arbitrary initial amplitude
distributions, while still maintaining $O(\sqrt{N})$ complexity.


\subsubsection{Inversion about the Average}
\label{inversion}

To perform inversion\index{inversion about average} about the average on 
a quantum computer the inversion must be a
unitary transformation\index{unitary transformation}.  Furthermore, 
in order for the algorithm as a whole to solve the problem
in $O(\sqrt N)$ time, the inversion must be able to be performed efficiently.
As will be shown shortly, the inversion can be accomplished 
with $O(n)=O(\log(N))$ quantum gates.

It is easy to see that the transformation 
\begin{displaymath}
\sum_{i=0}^{N-1}a_i\ket{x_i} \to \sum_{i=0}^{N-1}(2A-a_i)\ket{x_i},
\end{displaymath}
where $A$ denotes the average of the $a_i$'s, 
is performed by the $N\times N$ matrix
\begin{displaymath}
D = \left(\begin{array}{cccc} {2\over N}-1 & {2\over N} & \dots & {2\over N}\\
				{2\over N} & {2\over N}-1 & \dots & {2\over N}\\
				\dots & \dots & \dots & \dots\\
				{2\over N} & {2\over N} & \dots & {2\over N}-1\end{array}\right).
\end{displaymath}

Since $DD^* = I$, $D$ is unitary  and is therefore a 
possible quantum state transformation.  

We now turn to the question of how efficiently the transformation
can be performed, and show that it can be decomposed into 
$O(n) = O(\log(N))$ elementary quantum gates. 
Following Grover\index{Grover}, $D$ can be defined as $D = WRW$ where $W$ is the Walsh-Hadamard 
transform defined in section \ref{gates} and 
\begin{displaymath}
R = \left(\begin{array}{cccc}1 & 0 & \dots & 0\\
			       0 & -1 & 0 & \dots\\
			       0 & \dots & \dots & 0\\
			       0 & \dots & 0 & -1 \\\end{array}\right).
\end{displaymath}

To see that $D = WRW$, consider $R = R' - I$ where $I$ is the identity and
\begin{displaymath}
R' =\left(\begin{array}{cccc}2 & 0 & \dots & 0\\
			       0 & 0 & 0 & \dots\\
			       0 & \dots & \dots & 0\\
			       0 & \dots & 0 & 0 \\\end{array}\right).
\end{displaymath}
Now $WRW = W (R' - I) W = WR'W - I$.  It is easily verified that
\begin{displaymath}
WR'W = \left(\begin{array}{cccc}{2\over N} &{2\over N}  & \dots & {2\over N}\\
			       {2\over N} & {2\over N} & {2\over N} & \dots\\
			       {2\over N} & \dots & \dots & {2\over N}\\
			       {2\over N} & \dots & {2\over N} & {2\over N} \\\end{array}\right)
\end{displaymath}
and thus $WR'W - I = D$.

\subsubsection{Changing the Sign}
\label{ChangingSigns}

We still have to explain how to invert the amplitude of the 
desired result. We show, more generally, a surprising simple
way to invert the amplitude of exactly those states with
$P(x)=1$ for a general $P$.

Let $U_P$ be the gate array that performs the 
computation $U_P: \ket {x, b} \to \ket{x, b\oplus P(x)}$.  
Apply $U_P$ to the superposition 
$\ket{\psi} = {1\over \sqrt{2^n}}\sum_{x=0}^{n-1}\ket{x}$ and
choose $b = {1\over \sqrt 2}\ket 0 - \ket 1$ to end up in a state where the
sign of all $x$ with $P(x) = 1$ has been changed, and $b$ is
unchanged.  

To see this, let 
$X_0=\{x|P(x)=0\}$ and $X_1=\{x|P(x)=1\}$ and consider the application of $U_P$.
\begin{eqnarray*}
\lefteqn{U_P(\ket{\psi,b})}\\
&=& {1\over \sqrt{2^{n+1}}}U_P(\sum_{x\in X_0}\ket{x,0}+\sum_{x\in X_1}\ket{x,0} - \sum_{x\in X_0}\ket{x,1}-\sum_{x\in X_1}\ket{x,1})\\
	&=& {1\over \sqrt{2^{n+1}}}(\sum_{x\in X_0}\ket{x,0\oplus 0}+\sum_{x\in X_1}\ket{x,0\oplus 1} - \sum_{x\in X_0}\ket{x,1\oplus 0}-\sum_{x\in X_1}\ket{x,1 \oplus 1})\\
	&=& {1\over \sqrt{2^{n+1}}}(\sum_{x\in X_0}\ket{x,0}+\sum_{x\in X_1}\ket{x,1} - \sum_{x\in X_0}\ket{x,1}-\sum_{x\in X_1}\ket{x,0})\\
	&=& {1\over \sqrt{2^{n}}}(\sum_{x\in X_0}\ket{x} - \sum_{x\in X_1}\ket{x})\otimes b\\
\end{eqnarray*}
Thus the amplitude of the states in $X_1$ have been inverted as desired.

\subsection{Heuristic Search}

\subsubsection{A Note on the Walsh-Hadamard Transform}
\label{walsh-set}

There is another representation for the Walsh-Hadamard
transformation of section \ref{Walsh} that is useful for 
understanding how to use the Walsh-Hadamard transformation
in constructing quantum algorithms. The $n$ bit 
Walsh-Hadamard transformation is a 
$2^n\times 2^n$ matrix $W$ with entries $W_{rs}$ where both $r$ and
$s$ range from $0$ to $2^n-1$. We will show that 
\begin{displaymath}
W_{rs}=\frac{1}{\sqrt{2^n}}(-1)^{r\cdot s}
\end{displaymath}
where $r\cdot s$ is the number of common $1$ bits in the 
the binary representations of $r$ and $s$.

To see this equality, note that 
\begin{displaymath}
W(\ket{r})=\sum_sW_{rs}\ket{s}.
\end{displaymath}
Let $r_{n-1}\dots r_0$ be the binary representation of $r$, and
$s_{n-1}\dots s_0$ be the binary representation of $s$.
\begin{eqnarray*}
W(\ket{r})&=&(H\otimes\dots\otimes H)(\ket{r_{n-1}}\otimes\dots\otimes\ket{r_0})\\
        &=&\frac{1}{\sqrt{2^n}}(\ket{0}+(-1)^{r_{n-1}}\ket{1})\otimes\dots\otimes(\ket{0}+(-1)^{r_{0}}\ket{1})\\
       &=&\frac{1}{\sqrt{2^n}}\sum_{s=0}^{2^n-1}(-1)^{s_{n-1}r_{n-1}}\ket{s_{n-1}}\otimes\dots\otimes(-1)^{s_{0}r_{0}}\ket{s_{0}}\\
      &=&\frac{1}{\sqrt{2^n}}\sum_{s=0}^{2^n-1}(-1)^{s\cdot r}\ket{s}.
\end{eqnarray*}

\subsubsection{Overview of Hogg's algorithms}

\begin{figure}[t]
\setlength{\unitlength}{0.8pt}
\begin{picture}(396,320)

\put(192,0){$\emptyset$}

\put(72,59){$\{v_1=0\}$}
\put(144,59){$\{v_1=1\}$}
\put(216,59){$\{v_2=0\}$}
\put(288,59){$\{v_2=1\}$}

\put(0,130){$\left\{ \begin{array}{l}v_2=0\\v_2=1\end{array}\right\}$}
\put(72,130){$\left\{ \begin{array}{l}v_1=1\\v_2=1\end{array}\right\}$}
\put(144,130){$\left\{ \begin{array}{l}v_1=0\\v_2=1\end{array}\right\}$}
\put(216,130){$\left\{ \begin{array}{l}v_1=1\\v_2=0\end{array}\right\}$}
\put(288,130){$\left\{ \begin{array}{l}v_1=0\\v_2=0\end{array}\right\}$}
\put(360,130){$\left\{ \begin{array}{l}v_1=0\\v_1=1\end{array}\right\}$}

\put(62,210){$\left\{ \begin{array}{l}v_1=1\\v_2=0\\v_2=1\end{array}\right\}$}
\put(134,210){$\left\{ \begin{array}{l}v_1=0\\v_2=0\\v_2=1\end{array}\right\}$}
\put(206,210){$\left\{ \begin{array}{l}v_1=0\\v_1=1\\v_2=1\end{array}\right\}$}
\put(278,210){$\left\{ \begin{array}{l}v_1=0\\v_1=1\\v_2=0\end{array}\right\}$}

\put(170,305){$\left\{ \begin{array}{l}v_1=0\\v_1=1\\v_2=0\\v_2=1\end{array}\right\}$}

\put(195,17){\line(-3,1){108}}
\put(195,17){\line(-1,1){36}}
\put(195,17){\line(1,1){36}}
\put(195,17){\line(3,1){108}}

\put(87,76){\line(-2,1){72}}
\put(87,76){\line(0,1){36}}
\put(87,76){\line(2,1){72}}

\put(159,76){\line(-4,1){144}}
\put(159,76){\line(2,1){72}}
\put(159,76){\line(4,1){144}}

\put(231,76){\line(-4,1){144}}
\put(231,76){\line(0,1){36}}
\put(231,76){\line(4,1){144}}

\put(303,76){\line(-4,1){144}}
\put(303,76){\line(0,1){36}}
\put(303,76){\line(2,1){72}}

\put(15,150){\line(2,1){72}}
\put(15,150){\line(4,1){144}}

\put(87,150){\line(0,1){36}}
\put(87,150){\line(4,1){144}}

\put(159,150){\line(0,1){36}}
\put(159,150){\line(2,1){72}}

\put(231,150){\line(-4,1){144}}
\put(231,150){\line(2,1){72}}

\put(303,150){\line(-4,1){144}}
\put(303,150){\line(0,1){36}}

\put(375,150){\line(-4,1){144}}
\put(375,150){\line(-2,1){72}}

\put(87,238){\line(3,1){108}}
\put(159,238){\line(1,1){36}}
\put(231,238){\line(-1,1){36}}
\put(303,238){\line(-3,1){108}}
\end{picture}
\setlength{\unitlength}{1em}

\caption{Lattice of variable assignments in a CSP}
\label{setlat}
\end{figure}
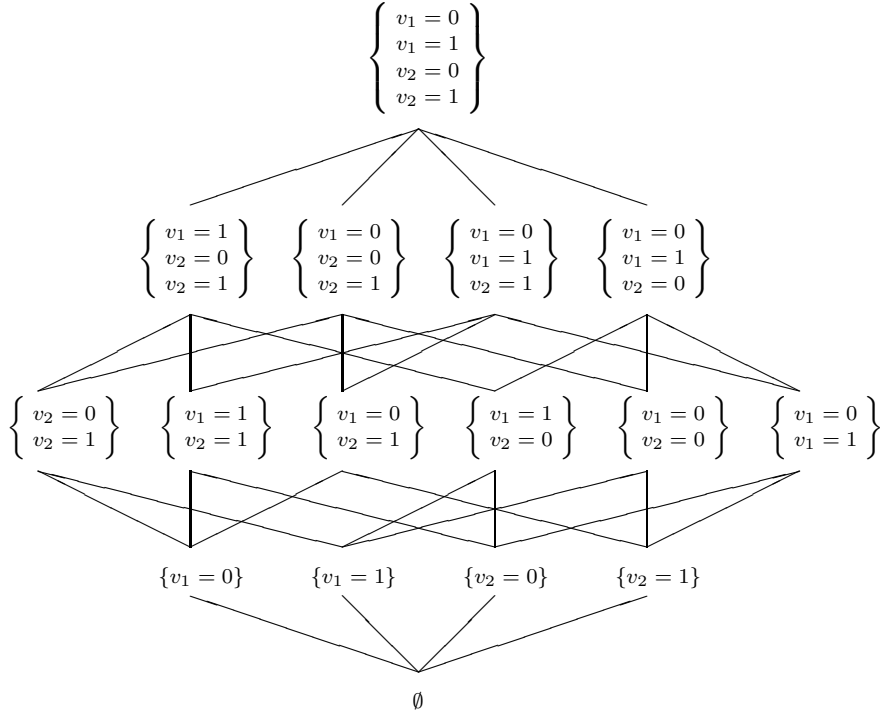

A constraint satisfaction problem (CSP) has $n$ variables $V=\{v_1,\dots,v_n\}$
which can take $m$ different values $X=\{x_1,\dots,x_m\}$ subject to certain
constraints $C_1,\dots,C_l$.  Solutions to a constraint satisfaction problem
lie in the space of assignments of $x_i$'s to $v_j$'s, $V\times X$. There is a 
natural lattice structure on this space given by set containment.  Figure \ref{setlat}
shows the assignment space and its lattice structure for $n=2$,
$m=2$, $x_1=0$, and $x_2=1$.
Note that the lattice includes both incomplete and inconsistent
assignments.

\begin{figure}[t]
\setlength{\unitlength}{0.8pt}
\begin{picture}(396,240)

\put(180,0){$\ket{0000}$}

\put(72,54){$\ket{1000}$}
\put(144,54){$\ket{0100}$}
\put(216,54){$\ket{0010}$}
\put(288,54){$\ket{0001}$}

\put(0,108){$\ket{1100}$}
\put(72,108){$\ket{1010}$}
\put(144,108){$\ket{1001}$}
\put(216,108){$\ket{0110}$}
\put(288,108){$\ket{0101}$}
\put(360,108){$\ket{0011}$}

\put(72,162){$\ket{1110}$}
\put(144,162){$\ket{1101}$}
\put(216,162){$\ket{1011}$}
\put(288,162){$\ket{0111}$}

\put(180,216){$\ket{1111}$}

\put(195,15){\line(-3,1){108}}
\put(195,15){\line(-1,1){36}}
\put(195,15){\line(1,1){36}}
\put(195,15){\line(3,1){108}}

\put(87,69){\line(-2,1){72}}
\put(87,69){\line(0,1){36}}
\put(87,69){\line(2,1){72}}

\put(159,69){\line(-4,1){144}}
\put(159,69){\line(2,1){72}}
\put(159,69){\line(4,1){144}}

\put(231,69){\line(-4,1){144}}
\put(231,69){\line(0,1){36}}
\put(231,69){\line(4,1){144}}

\put(303,69){\line(-4,1){144}}
\put(303,69){\line(0,1){36}}
\put(303,69){\line(2,1){72}}

\put(15,123){\line(2,1){72}}
\put(15,123){\line(4,1){144}}

\put(87,123){\line(0,1){36}}
\put(87,123){\line(4,1){144}}

\put(159,123){\line(0,1){36}}
\put(159,123){\line(2,1){72}}

\put(231,123){\line(-4,1){144}}
\put(231,123){\line(2,1){72}}

\put(303,123){\line(-4,1){144}}
\put(303,123){\line(0,1){36}}

\put(375,123){\line(-4,1){144}}
\put(375,123){\line(-2,1){72}}

\put(87,177){\line(3,1){108}}
\put(159,177){\line(1,1){36}}
\put(231,177){\line(-1,1){36}}
\put(303,177){\line(-3,1){108}}
\end{picture}
\setlength{\unitlength}{1em}

\caption{Lattice of variable assignments in ket form}
\label{ketlat}
\end{figure}
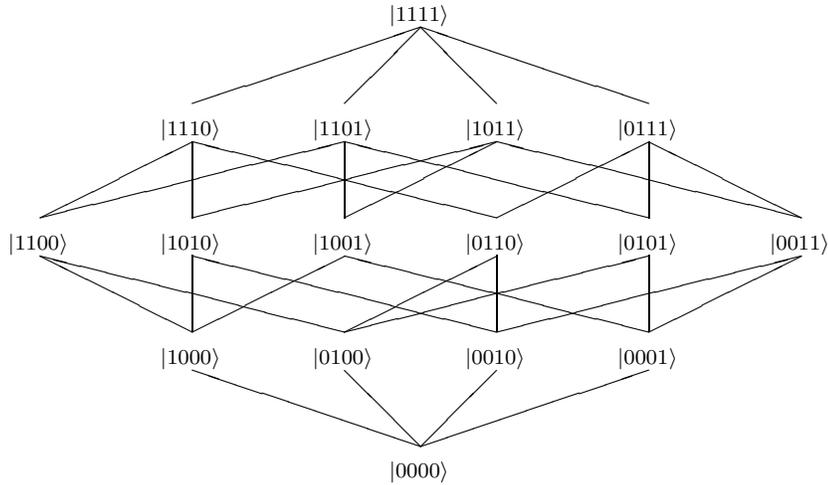

Using the standard correspondence between sets of enumerated elements and binary
sequences, in which a $1$ in the $n$th place corresponds to inclusion of
the $n$th element and a $0$ corresponds to exclusion, standard
basis vectors for a quantum state space can be put
in one to one correspondence with the sets. For example, Figure \ref{ketlat} shows the
lattice of Figure \ref{setlat} rewritten in ket notation
where the elements $v_1=0$, $v_1=1$, $v_2=0$ and $v_2=1$ have been
enumerated in that order.

If a state violates a constraint, then so do all
states above it in the lattice.  
The approach Hogg takes in designing quantum algorithms for constraint
satisfaction problems is to begin with all the amplitude concentrated
in the $\ket{0\dots 0}$ state and to iteratively
move amplitude up the lattice from sets to supersets and away from sets
that violate the constraints. Note that this algorithm begins differently 
than Shor's algorithm and Grover's algorithm, which both begin by
computing a function on a superposition of all the input values at once.

Hogg gives two ways \cite{Hogg-96,Hogg-98} of constructing 
a unitary matrix for moving
amplitude up the lattice. We will describe both methods, and then
describe how he moves amplitude away from bad sets.

{\bf Moving amplitude up: Method 1.} There is an obvious transformation
that moves amplitude from sets to 
supersets. Any amplitude associated to the empty set is evenly
distributed among all sets with a single element. Any amplitude
associated to a set with a single element is evenly distributed
among all two element sets which contain that element and so on.  For
the lattice of a three element set 
\begin{displaymath}
\begin{array}{ccccc}
        &       &\ket{111}&     &       \\
        &\lup   &\lvert &\ldown &       \\
\ket{011}&      &\ket{101}&     &\ket{110}\\
\lvert  &\lcross&       &\lcross&\lvert \\
\ket{001}&       &\ket{010}&     &\ket{100}\\
        &\ldown &\lvert &\lup   &       \\
        &       &\ket{000}&     &       \\
\end{array}
\end{displaymath}
We want to transform
\begin{eqnarray*}
\ket{000} & \to & 1/\sqrt{3} (\ket{001} + \ket{010} + \ket{100}\\
\ket{001} & \to & 1/\sqrt{3} (\ket{011} + \ket{110} + \ket{101}\\
&\dots&\\
\end{eqnarray*}
The complete matrix for this transformation looks like (as usual the basis vectors
are ordered according to their binary representation)
\begin{displaymath}
 \left(
\begin{array}{cccccccc}
0&0&0&0&0&0&0&1\\
\frac{1}{\sqrt 3}&0&0&0&0&0&0&0\\
\frac{1}{\sqrt 3}&0&0&0&0&0&0&0\\
0&\frac{1}{\sqrt 2}&\frac{1}{\sqrt 2}&0&0&0&0&0\\
\frac{1}{\sqrt 3}&0&0&0&0&0&0&0\\
0&\frac{1}{\sqrt 2}&0&0&\frac{1}{\sqrt 2}&0&0&0\\
0&0&\frac{1}{\sqrt 2}&0&\frac{1}{\sqrt 2}&0&0&0\\
0&0&0&1&0&1&1&0
\end{array}
\right)
\end{displaymath}

Unfortunately this transformation is not unitary. Hogg \cite{Hogg-96} 
uses the fact that the
closest (in a suitable metric) unitary matrix $U_M$ to an arbitrary matrix $M$ can
be found using $M$'s singular value decomposition $M=UDV^T$ where
$D$ is a diagonal matrix, and $U$ and $V$ are unitary matrices. The
product $U_M=UV^T$ gives the closest unitary matrix to $M$. Provided
that $U_M$ is sufficiently close to $M$, 
$U_M$ will behave in a similar way to $M$ and will therefore
do a reasonably job of moving amplitude from sets to their supersets.

{\bf Moving amplitude up: Method 2.}The second approach \cite{Hogg-98} uses the Walsh-Hadamard transformation. Hogg
assumes that the desired matrix has form $WDW$ where $W$ is the
Walsh-Hadamard transformation and $D$ is a diagonal matrix
whose entries depend only on the size of the sets. Hogg 
calculates the entries for $D$ which maximize the movement
of amplitude from a set to its supersets. 
This calculation exploits the property
\begin{displaymath}
W_{rs}=\frac{1}{\sqrt{N}}(-1)^{|r\cdot s|}=\frac{1}{\sqrt{N}}(-1)^{|r\cap s|}
\end{displaymath}
shown in section \ref{walsh-set}.

{\bf Moving amplitude away from bad sets.} To effect moving amplitude away from sets that violate the constraints,
Hogg suggests adjusting the phases of the sets, depending on the
extent to which they violate the constraints, in such a way that
amplitude distributed to sets that have bad subsets cancels, where
as the amplitude distributed to sets from all good subsets adds.
Different choices here will work more
or less effectively depending on the particular problem. One
choice he suggests is inverting the phase of all bad sets which
will result in some cancelation in the amplitude of supersets between
the amplitude coming from good subsets and bad subsets.
This phase inversion can be done as in Grover's algorithm
(\ref{ChangingSigns}) with a 
$P$ that tests whether a given state satisfies all of the constraints or not. 
Another suggestion is to give random phases to the bad
sets so that on average the contribution to the amplitude of a 
superset from bad subsets is zero. Other choices are possible.

Because the canceling resulting from the phase changes varies from
problem to problem, the probability of obtaining a solution is
difficult to analyse. A few small experiments have been done and the
guess is that the cost of the search still grows exponentially, but
considerably more slowly than in the unstructured case.
But until sufficiently large quantum computers are
built, or better techniques for analyzing such algorithms are found,
the efficiency cannot be determined for sure.

\section{Quantum Error Correction}

\label{qec}

One fundamental problem in building quantum computers is the need to isolate
the quantum state.  An interaction of particles representing
qubits with the external environment 
disturbs the quantum state, and causes it to decohere\index{decoherence}, 
or transform in an unintended and often non-unitary fashion.  

Steane\index{Steane} \cite{Steane-97} estimates that
the decoherence\index{decoherence} 
of any system likely to be built is $10^7$ times too large to be able to run
Shor's algorithm\index{Shor's algorithm} as it stands on a $130$ digit number.
However, adding error correction algorithms to Shor's algorithm mitigates the
effect of decoherence, making it again look possible that a system could
be built on which Shor's algorithm
could be run for large numbers. 

On the surface quantum error correction is similar to classical error
correcting codes in that redundant bits are used to detect and correct
errors.  But the situation for quantum error correction is somewhat 
more complicated than in the classical case since we are not dealing
with binary data but with quantum states.  

Quantum error correction\index{quantum error correction} must reconstruct
the exact encoded quantum state. Given the impossibility of cloning or 
copying the quantum state, this reconstruction appears harder than
in the classical case. However, it turns out that classical techniques 
can be modified to work for quantum systems.

\subsection{Characterization of Errors}

In the following it is assumed that all errors are the result of quantum 
interaction between a set of qubits and the environment.  
The possible errors for each single qubit considered are linear combinations of
no errors ($I$), bit flip errors ($X$), phase errors ($Z$), and bit flip phase
errors ($Y$).  A general single bit error is thus a transformation 
$e_1 I + e_2 X + e_3 Y + e_4 Z$.  
Interaction with the environment transforms single qubits according to
\begin{displaymath}
\ket{\psi} \to (e_1 I + e_2 X + e_3 Y + e_4 Z)\ket{\psi} = \sum_i{e_iE_i\ket{\psi}}.
\end{displaymath}

For the general case of quantum registers, possible errors are expressed as linear combinations
of unitary error operators $E_i$.  These could be 
combinations of single bit errors, like
tensor products of the single bit error transformations $\{I, X, Y, Z\}$, or more general
multi-bit transformations.  In any case, an error can be written as 
$\sum_i{e_iE_i}$ for some error operators $E_i$ and coefficients $e_i$.

\subsection{Recovery of Quantum State}

An error correcting code for a set of errors $E_i$ consists of a
mapping $C$ that embeds $n$ data bits in $n+k$ code bits together with
a syndrome extraction operators $S_C$ that maps $n+k$ code bits to the 
set of indices of correctable errors $E_i$ such that 
$i = S_C (E_i (C (x)))$.  If $y = E_j (C(x))$ for some unknown but 
correctable error, then error $S_C(y)$ can be used to recover a 
properly encoded value $C(x)$, i.e.~$E_{S_C(y)}^{-1}(y) = C(x)$.

Now consider the case of a quantum register.  First, the state of the
register can be in a superposition of basis vectors.  Furthermore, the
error can be a combination of correctable error operators $E_i$.  It turns out that
it is still possible to recover the encoded quantum state. 

Given an error correcting code $C$ with syndrome extraction 
operator\index{syndrome extraction operator} $S_C$, an
$n$-bit quantum state $\ket {\psi}$ is encoded in a $n+k$ bit quantum state
$\ket{\phi} = C \ket{\psi}$.
Assume that decoherence\index{decoherence} leads to an error state $\sum_i{e_iE_i\ket{\phi}}$ for
some combination of correctable errors $E_i$.  The original encoded state $\ket{\phi}$
can be recovered as follows:

\begin{enumerate}
\item Apply the syndrome extraction operator $S_C$ to the quantum state padded with
sufficient $\ket 0$ bits:
\begin{displaymath}
S_C (\sum_i{e_iE_i\ket{\phi}}) \otimes \ket 0 = \sum_i{e_i(E_i\ket{\phi}\otimes \ket i)}.
\end{displaymath}
Quantum parallelism\index{quantum parallelism} gives a superposition of different errors each associated with
their respective error index $i$.
\item Measure the $\ket i$ component of the result.  This yields some (random) value $i_0$ and
projects the state to 
\begin{displaymath}
E_{i_0}\ket{\phi,i_0}
\end{displaymath}

\item Apply the inverse error transformation $E_{i_0}^{-1}$ to the 
first $n+k$ qubits of $E_{i_0}\ket{\phi,i_0}$ to
get the corrected state $\ket{\phi}$.
\end{enumerate}

Note that step 2 projects a superposition of multiple error transformations into 
a single error.  Consequently, only one inverse error transformation is required 
in step 3.

\subsection{Error Correction Example}

Consider the trivial error correcting code $C$ that 
maps $\ket 0 \to \ket {000}$ and $\ket 1 \to \ket{111}$.  $C$ can correct single bit
flip errors 
\begin{displaymath}
E = \{I\otimes I\otimes I, X\otimes I\otimes I, I\otimes X\otimes I, I\otimes I\otimes X\}.
\end{displaymath}

The syndrome extraction operator is
\begin{displaymath}
S:  \ket {x_0,x_1,x_2,0, 0, 0} \to \ket {x_0,x_1,x_2,x_0\xor x_1,x_0\xor x_2,x_1\xor x_2},
\end{displaymath}
with the corresponding error correction operators shown in the table. 
Note that $E_i = E_i^{-1}$ for this example.
\begin{center}
\begin{tabular}{r|c|c}
Bit flipped & Syndrome & Error correction \\\hline
none	& $\ket {000}$ & none \\
$0$	& $\ket {110}$ & $X\otimes I\otimes I$\\
$1$	& $\ket {101}$ & $I\otimes X\otimes I$\\
$2$	& $\ket {011}$ & $I\otimes I\otimes X$\\
\end{tabular}
\end{center}

Consider the quantum bit $\ket {\psi} = {1 \over \sqrt 2} (\ket 0 - \ket 1) $ that is encoded as
\begin{displaymath}
C\ket{\psi} = \ket{\phi} = {1 \over \sqrt 2} (\ket {000} - \ket {111})
\end{displaymath}
and the error \begin{displaymath}
E = {4 \over 5} X\otimes I \otimes I + {3 \over 5} I\otimes X \otimes I.
	      \end{displaymath}
The resulting error state is 
\begin{eqnarray*}
E \ket{\phi} &=& 
	({4 \over 5} X\otimes I \otimes I + {3 \over 5} I\otimes X \otimes I)({1 \over \sqrt 2} (\ket {000} - \ket {111}))\\
	&=& {4 \over 5} X\otimes I \otimes I({1 \over \sqrt 2} (\ket {000} - \ket {111})) + {3 \over 5} I\otimes X \otimes I({1 \over \sqrt 2} (\ket {000} - \ket {111}))\\
	&=& {4 \over 5\sqrt 2} X\otimes I \otimes I (\ket {000} - \ket {111}) + {3 \over 5\sqrt 2} I\otimes X \otimes I (\ket {000} - \ket {111})\\
	&=& {4 \over 5\sqrt 2} (\ket {100} - \ket {011}) + {3 \over 5\sqrt 2} (\ket {010} - \ket {101})\\
\end{eqnarray*}
	
Next apply the syndrome extraction to $(E\ket{\phi})\otimes \ket {000}$ as follows:
\begin{eqnarray*}
\lefteqn{S_C((E\ket{\phi})\otimes \ket {000})} \\
&=&S_C({4 \over 5\sqrt 2} (\ket {100000} - \ket {011000}) + {3 \over 5\sqrt 2} (\ket {010000} - \ket {101000}))\\
&=&{4 \over 5\sqrt 2} (\ket {100110} - \ket {011110}) + {3 \over 5\sqrt 2} (\ket {010101} - \ket {101101})\\
&=&{4 \over 5\sqrt 2} (\ket {100} - \ket {011})\otimes \ket{110} + {3 \over 5\sqrt 2} (\ket {010} - \ket {101})\otimes \ket{101}\\
\end{eqnarray*}
Measuring the last three bits of this state yields either $\ket{110}$ or $\ket{101}$.  
Assuming the measurement\index{measurement} produces the former, the state becomes 
\begin{displaymath}
{1 \over \sqrt 2} (\ket {100} - \ket {011})\otimes \ket{110}.
\end{displaymath}
 
The measurement has the almost magical effect of causing all but
one summand of the error to disappear. The remaining part of the error can
be removed by applying the inverse error operator
$X\otimes I\otimes I$, corresponding to the measured value $\ket{110}$,
to the first three bits, to produce 
\begin{displaymath}
{1 \over \sqrt 2} (\ket {000} - \ket {111}) = C\ket{\psi} = \ket{\phi}.
\end{displaymath}

\section{Conclusions}

Quantum computing is a new, emerging field that has the potential to
dramatically change the way we think about computation, programming
and complexity.  The challenge for computer scientists and others is
to develop new programming techniques appropriate for quantum
computers.  Quantum entanglement and phase cancellation 
introduce a new dimension to
computation.  Programming no longer consists of merely formulating step-by-step
algorithms but requires new techniques of adjusting phases, and mixing and diffusing
amplitudes to extract useful output.

We have tried to give an accurate account of the state-of-the-art of quantum 
computing for computer scientists and other non-physicists.  We have described
some of the quantum mechanical effects, like  the exponential state 
space, the entangled states, and the linearity of quantum state transformations, that
make quantum parallelism possible.  Even though quantum computations must be 
linear and reversible, any classical algorithm can be implemented on a quantum
computer.  
But the real power of these new machines, the exponential parallelism,
can only be exploited using new, innovative programming techniques.
People have only recently begun to research such techniques.  

We have
described Shor's polynomial-time factorization algorithm that 
stimulated the field of quantum computing.  Given a practical quantum
computer, Shor's algorithm would make many present cryptographic
methods obsolete.  Grover's search algorithm, while only providing 
a polynomial speed-up, proves that
quantum computers are strictly more powerful than classical ones.  Even though
Grover's algorithm has been shown to be optimal, there is hope 
that faster algorithms can be found by exploiting properties of 
the problem structure.  We have described
one such approach taken by Hogg.  

There are a few other known quantum algorithms that we did
not discuss.  Jones and Mosca \cite{Jones+Mosca-98} describe the
implementation on a 2-bit quantum computer of a constant
time algorithm \cite{Deutsch-Jozsa-91} that can
distinguish whether a function is balanced or constant.  Grover
\cite{Grover-xxx} describes an efficient algorithm for estimating the
median of a set of values and both Grover \cite{Grover-xxx} and
Terhal and Smolin \cite{Terhal+Smolin-97} 
using different methods can solve the coin weighing problem in a
single step.

Beyond these algorithms not much more is known about what could be done with a 
practical quantum computer.  It is an open question whether or not
we can find quantum algorithms that provide exponential speed-up for problems
other than factoring.
There is some speculation among physicists that quantum 
transformations might be slightly non-linear. So far all
experiments that have been done are consistent with the standard linear
quantum mechanics, but a slight non-linearity
is still possible.  Abrams\index{Abrams} and Lloyd\index{Lloyd}
\cite{AbramsLloyd98} show that even a very slight non-linearity
could be exploited to solve all NP hard problems on a quantum
computer in polynomial time. This result further highlights the
fact that computation is fundamentally a physical process, and that 
what can be computed efficiently may depend on subtle issues in physics.

The unique properties of quantum computers give rise to new kinds of
complexity classes.  For instance, BQP is the set of all languages
accepted by a quantum Turing machine in polynomial time with bounded
probability.  Details of the extensive research done in the field of
quantum complexity theory is beyond the scope of this paper.  The
interested reader may start by consulting \cite{Bennett-et-al-97} and
\cite{Watrous98} respectively for analyses of time and space
complexity of quantum computation.  \cite{Williams-98} contains an
introduction to early results in quantum complexity.

Of course, there are daunting physical problems that must be overcome if
anyone is ever to build a useful quantum computer. Decoherence, the
distortion of the quantum state due to interaction with the environment,
is a key problem.  A big breakthrough for dealing with decoherence came
from the algorithmic, rather than the physical, side of the field with the
development of quantum error correction techniques.  We have described
some of the principles involved.  Further advances in quantum error
correction and the development of robust algorithms will be as important
for the development of practical quantum computers 
as advances in the hardware side. 

\subsection{Further Reading}

Andrew Steane's survey article ``Quantum computing'' \cite{Steane-97} 
is aimed at physicists. 
We recommend reading his paper for his viewpoint on this subject,
particularly for his description of connections between information
theory and quantum computing and for his discussion of error correction,
of which he was one of the main developers. He also has an overview of
the physics involved in actually building quantum computers, and a survey
of what had been done up to July 1997. His article contains a more 
detailed history of the ideas related to quantum computing than the
present paper, and has more references as well.
Another shorter and very readable tutorial can be found in \cite{Berthiaume97}.

Richard Feynman's {\em Lectures on Computation} \cite{Feynman-96}
contains a reprint of the lecture ``Quantum Mechanical Computers''
\cite{Feynman-85} which began the whole field. It also discusses
the thermodynamics of computations which is closely tied with
reversible computing and information theory.

Colin Williams and Scott Clearwater's book {\em Explorations in
Quantum Computing} \cite{Williams-98} comes with software, in the
form of Mathematica notebooks, that simulates some
quantum algorithms like Shor's algorithm.

The second half of the October 1997 issue of the SIAM Journal
of Computing contains six seminal articles on quantum computing,
including four we have already cited 
\cite{Bennett-et-al-97} \cite{Bernstein-Vazirani-93} \cite{Shor-95} \cite{Simon-94}.

Most of the articles referenced in this paper, and many more,
can be found at the Los Alamos preprint server:
{\url http://xxx.lanl.gov/archive/quant-ph}.
Links to research projects and other information about quantum computing can
be found on our web site {\url http://www.pocs.com/qc.html}.

\bibliographystyle{esub2acm}
\bibliography{qc}

\begin{acks}

The authors would like to thank Tad Hogg and Carlos Mochon for
many enjoyable conversations about quantum computing, and for their
feedback on an earlier draft of this paper. We are also grateful to 
Lee Corbin, David Goldberg, Lov Grover, Norman Hardy, Vaughan Pratt, 
Marc Rieffel and the anonymous referees 
for detailed comments on earlier drafts of this paper. 
Finally, we would
like to thank FXPAL for enthusiastically supporting this work.

\end{acks}

\appendix
\section{Tensor Products}

\label{tensor-product}

The tensor product\index{tensor product} ($\otimes$) of a $n$-dimensional
and a $k$-dimensional vector is a $nk$-dimensional
vector.  Similarly, if $A$ and $B$ are transformations on 
$n$-dimensional and a $k$-dimensional vectors respectively, then $A\otimes
B$\footnote{Technically, this is a right Kronecker product.} is a
transformation on $nk$-dimensional vectors.

The exact mathematical details of tensor products are beyond the scope
of this paper (see \cite{Hungerford-74} for a comprehensive treatment).
For our purposes the following algebraic rules are sufficient to
calculate with tensor products.  For matrices $A$,$B$,$C$,$D$, $U$, vectors
$u$, $x$, $y$, and scalars $a$, $b$ the following hold:

\begin{eqnarray*}
(A \otimes B) (C \otimes D) &=& AC\otimes BD\\
(A \otimes B) (x \otimes y) &=& Ax\otimes By\\
(x+y)\otimes u&=& x\otimes u + y\otimes u\\
u\otimes(x+y)&=& u\otimes x + u\otimes y\\
ax\otimes by &=& ab(x\otimes y)
\end{eqnarray*}

\begin{displaymath}
\left(\begin{array}{cc}A & B\\C & D\end{array}\right) \otimes U = 
\left(\begin{array}{cc}A \otimes U & B \otimes U\\C \otimes U & D \otimes U\end{array}\right),
\end{displaymath}
which specialized for scalars $a, b, c, d$ to
\begin{displaymath}
\left(\begin{array}{cc}a & b\\c & d\end{array}\right) \otimes U = 
\left(\begin{array}{cc}a U & b U\\c U & d U\end{array}\right).
\end{displaymath}

The conjugate transpose\index{conjugate transpose} distributes over tensor products, i.e.
\begin{displaymath}
(A\otimes B)^*= A^*\otimes B^*.
\end{displaymath}

A matrix $U$ is {\em unitary} \index{unitary} if its conjugate transpose\index{conjugate transpose} its inverse:
$U^*U=I$.

The tensor product\index{tensor product} of several matrices is unitary if and only if each one of the
matrices is unitary up to a constant.  Let $U = A_1\otimes A_2\otimes \dots \otimes A_n$.  Then
$U$ is unitary if $A_i^*A_i = k_i I$ and $\Pi_ik_i = 1$.
\begin{eqnarray*}U^*U &=& (A_1^*\otimes A_2^*\otimes \dots \otimes A_n^*)(A_1\otimes A_2\otimes \dots \otimes A_n)\\
&=& A_1^*A_1\otimes A_2^*A_2\otimes \dots \otimes A_n^*A_n\\
&=& k_1I\otimes \dots k_nI\\
&=& I\\
\end{eqnarray*}
where each $I$ refers to the identity matrix of appropriate dimension.

For example, the distributive law\index{distributive law} allows computations of the form:
\begin{eqnarray*}
&&(a_0\ket 0 + b_0\ket 1) \otimes (a_1\ket 0 + b_1\ket 1)\\
		    &=& (a_0\ket 0 \otimes a_1\ket 0) + (b_0\ket 1 \otimes a_1\ket 0) +
			(a_0\ket 0 \otimes b_1\ket 1) + (b_0\ket 1 \otimes b_1\ket 1)\\
		    &=& a_0a_1((\ket 0 \otimes \ket 0) + b_0a_1(\ket 1 \otimes \ket 0) +
			a_0b_1(\ket 0 \otimes \ket 1) + b_0b_1(\ket 1 \otimes \ket 1)\\
		    &=& a_0a_1(\ket {00} + b_0a_1\ket {10} +
			a_0b_1\ket {01} + b_0b_1\ket {11}
\end{eqnarray*}

\section{Continued fractions and extracting the period from the measurement in Shor's algorithm}

\label{continued fractions}
In the general case where the period $r$ does not divide $2^m$,
the value $v$ measured in step 4 of Shor's algorithm
will be, 
with high probability, close to some multiple of $\frac{2^m}{r}$,
say $j\frac{2^m}{r}$.
 
The aim is to extract the period $r$ from the measured
value $v$. Shor shows that, with high probability, $v$ is within
$\frac{1}{2}$ of some $j\frac{2^m}{r}$. Thus
\begin{displaymath}
\left|v-j\frac{2^m}{r}\right| < \frac{1}{2}
\end{displaymath}
 for some $j$, which implies that
\begin{displaymath}
\left|\frac{v}{2^m}-\frac{j}{r}\right| < \frac{1}{2\cdot2^m} < \frac{1}{2M^2}.
\end{displaymath}
The difference between two distinct fractions $\frac{p}{q}$ and  $\frac{p'}{q'}$
with denominators less than $M$ is bounded 
\begin{displaymath}
\left|\frac{p}{q} - \frac{p'}{q'}\right| = \left|\frac{pq'-p'q}{qq'}\right| >
\frac{1}{M^2}.
\end{displaymath}
Thus there is at most one fraction  $\frac{p}{q}$ with denominator $q<M$
such that $\left|\frac{v}{2^m}-\frac{p}{q}\right|< \frac{1}{M^2}$.
In the high probability case that $v$ is within $\frac{1}{2}$ of
$j\frac{2^m}{r}$, this fraction will be $\frac{j}{r}$.
 
The unique
fraction with denominator less than M that is within $\frac{1}{M^2}$
of $\frac{v}{2^m}$ can be obtained efficiently from the continued fraction
expansion of $\frac{v}{2^m}$ as follows. 
Using the sequences
\begin{eqnarray*}
a_0&=&\left[\frac{v}{2^m}\right]\\
\epsilon_0&=&\frac{v}{2^m} - a_0\\
a_n&=&\left[\frac{1}{\epsilon_{n-1}}\right]\\
\epsilon_n &=& \frac{1}{\epsilon_{n-1}}-a_n\\
p_0&=&a_0\\
p_1&=&a_1a_0+1\\
p_n&=&a_n p_{n-1}+p_{n-2}\\
q_0&=&1\\
q_1&=&a_1\\
q_n&=&a_n q_{n-1}+q_{n-2}\\
\end{eqnarray*}
compute the first fraction $\frac{p_n}{q_n}$ such that $q_n < M \leq q_{n+1}$.
See any standard number theory text, like Hardy and Wright \cite{Hardy-Wright},
for why this procedure works.
 
In the high probability case when $\frac{v}{2^m}$ is within
$\frac{1}{M^2}$ of a multiple $\frac{j}{r}$ of $\frac{1}{r}$, the fraction
obtained from the above procedure is $\frac{j}{r}$ as it has
denominator less than $M$. 
We take the denominator $q$ of the obtained fraction as our
guess for the period, which will work 
when $j$ and $r$ are relatively prime. 

\end{document}